\documentstyle[12pt,jeep]{article}
\numberbysection
\date{}
\begin{document}
\title{Resonances, Radiation Damping  and Instability in Hamiltonian
Nonlinear Wave Equations}
\author{
A. Soffer \thanks{Department of Mathematics, Rutgers University, New
Brunswick, NJ} \hspace{.05 in}
 and 
 M.I. Weinstein \thanks{Department of Mathematics, 
University of Michigan, Ann Arbor, MI  
 }} 
\baselineskip=18pt
\maketitle
\newcommand{\ds}{\displaystyle}
\newcommand{\xtext}[1]{\mbox{#1}}
\newcommand{\ra}{\rightarrow}
\newcommand{\da}{\downarrow}
\newcommand{\wra}{\rightharpoonup}
\newcommand{\rf}{\widehat}
\newcommand{\nit}{\noindent}
\newcommand{\no}{\nonumber}
\newcommand{\be}{\begin{equation}}
\newcommand{\ee}{\end{equation}}
\newcommand{\ba}{\begin{eqnarray}}
\newcommand{\ea}{\end{eqnarray}}
\newcommand{\bom}{\mbox{$\Omega$}}
\newcommand{\Fee}{\mbox{$\Phi$}}
\newcommand{\del}{\mbox{$\nabla$}}
\newcommand{\delx}{\mbox{$\nabla_{x}$}}
\newcommand{\dely}{\mbox{$\nabla_{y}$}}
\newcommand{\ar}{\mbox{$\alpha$}}
\newcommand{\fee}{\mbox{$\varphi$}}
\newcommand{\dta}{\mbox{$\delta$}}
\newcommand{\lam}{\mbox{$\lambda$}}
\newcommand{\Lam}{\mbox{$\Lambda$}}
\newcommand{\eps}{\mbox{$\epsilon$}}
\newcommand{\gam}{\mbox{$\gamma$}}
\newcommand{\Gam}{\mbox{$\Gamma$}}
\newcommand{\al}{\mbox{$\alpha$}}
\newcommand{\bt}{\mbox{$\beta$}}
\newcommand{\va}{\mbox{$\bf a $}}
\newcommand{\db}{\mbox{$\parallel$}}
\newcommand{\ov}{\overline}
\newcommand{\ud}{\underline}
\newcommand{\nW}{\left| W\right|_{\bf a}}
\newcommand{\D}{\partial}
\newcommand{\px}{\mbox{$\partial_{xx}^{2}$}}
\newcommand{\bu}{{\bf u}}
\newcommand{\bXz}{{\bf X}_0}
\newcommand{\bX}{{\bf X}}
\newtheorem{theo}{Theorem}[section]
\newtheorem{defin}{Definition}[section]
\newtheorem{prop}{Proposition}[section]
\newtheorem{lem}{Lemma}[section]
\newtheorem{cor}{Corollary}[section]
\newtheorem{rmk}{Remark}[section]
\renewcommand{\theequation}{\arabic{section}.\arabic{equation}}
\def\R{{\rm \rlap{I}\,\bf R}}
\def\C{{\bf C} \llap{
 \vrule height 6.6pt width 0.6pt depth 0pt \hskip 4.7pt}}
\def\WB{W\tilde g_\Delta(H)}
\def\tg{\tilde g_\Delta (H)}
\def\tA{\tilde A}
\def\Pc{{\bf P_c}}
\def\nn{\nonumber}
\def\veps{\varepsilon}
\newcommand{\nm}[1]{\vert\vert {#1} \vert\vert}
\def\la{\langle}
\def\ra{\rangle}
\def\jxs{\la x\ra^{\sigma}}
\def\jxms{\la x\ra^{-\sigma}}
\def\jt{\la t\ra}
\def\j#1{\la #1 \ra}
% local decay norm of \phi_d
\def\ld{[\phi_d]_{LD}}
\def\eW{\eta_W}
\def\Xnm#1{\nm{#1}_X}
\def\e#1{e^{#1}}
\def\i{\int^t_0}
\def\lab{\lambda\over 2iB}
\def\ov{\overline A}
\def\B{B+\Omega}
\def\de{\delta\downarrow 0}
\def\d{\delta}
\def\et{\eta^{nr}}
\def\va{\Pc\varphi^3}
\def\ezz{E_0^{(0)}(t)}
\def\eoz{E_1^{(0)}(t)}
\def\ez{E_0(t)}
\def\eo{E_1(t)}

%\centerline{\it To appear in Inventiones Mathematicae}
\medskip

\begin{abstract}
\nit 
We consider a class of nonlinear Klein-Gordon equations which are
Hamiltonian and are  
 perturbations of linear dispersive equations. The
unperturbed dynamical system has a  bound
state, a spatially localized and time periodic
solution. We show that, for generic nonlinear Hamiltonian
perturbations,  all small amplitude solutions decay to zero as time tends
to infinity at an anomalously slow rate. 
 In particular, spatially localized and 
 time-periodic solutions of the linear problem are
destroyed by generic nonlinear Hamiltonian perturbations via  
 slow radiation of energy to infinity. These solutions can therefore be
 thought of as {\it metastable states}.  
 The main mechanism  is a nonlinear resonant interaction
of bound states (eigenfunctions) and radiation (continuous spectral
modes), leading to energy transfer from the discrete to continuum
modes. 
 This is in contrast to the KAM theory in which appropriate 
 nonresonance conditions imply the persistence of invariant tori.
A hypothesis ensuring that such a resonance takes place is a
nonlinear analogue of the Fermi golden rule, arising in the theory of
resonances in quantum mechanics. The techniques used involve: (i) a
time-dependent method developed by the authors for the treatment of the quantum
resonance problem and perturbations of embedded eigenvalues, 
(ii) a generalization of the Hamiltonian normal
form appropriate for infinite dimensional dispersive systems and 
 (iii) ideas from scattering theory.
The arguments are quite general and we expect them to apply to a large
class of systems which can be viewed as the interaction of finite
dimensional and infinite dimensional dispersive dynamical systems, or as
a system of particles coupled to a field. 

\end{abstract}
\thispagestyle{empty}

\vfil\eject
\bigskip
\centerline {Table of Contents}
\medskip
\noindent{Section 1: Introduction and statement of main results - Smile
of the Cheshire Cat}

\noindent{Section 2: Linear analysis}

\noindent{Section 3: Existence theory}

\noindent{Section 4: 
Isolation of the key resonant terms and formulation
as a coupled finite and infinite dimensional dynamical  system}

\noindent{Section 5: Dispersive Hamiltonian normal form}

\noindent{Section 6: Asymptotic behavior of solutions of perturbed
normal form equations}

\noindent{Section 7: Asymptotic behavior of solutions of the nonlinear
Klein Gordon equation}

\noindent{Section 8: Summary and discussion}

\section{Introduction and statement of main results - Smile
of the Cheshire Cat$^*$}

Structural stability and 
the persistence of coherent structures under perturbations of a
dynamical system are  fundamental issues in dynamical
systems theory with implications in many fields of application.
In the context of discrete, finite dimensional Hamiltonian systems,
 this issue is addressed by the celebrated KAM theorem
\cite{kn:KAM}, which guarantees the persistence of most invariant tori
 of the unperturbed dynamics under small Hamiltonian perturbations.
For infinite dimensional systems, defined by Hamiltonian partial
differential equations (PDEs), KAM type methods have recently been used
to obtain results on the persistence of periodic and quasiperiodic
solutions, in the case where
solutions are defined on compact spatial domains with appropriate
boundary conditions \cite{kn:Bgn}, \cite{kn:CW}, \cite{kn:Kuk}.
 Variational methods have also been used to study this problem; see, 
 for example, \cite{kn:Bz}.
The compactness of the spatial domain ensures discreteness of the
spectrum associated with the unperturbed dynamics.  
  Therefore this situation is the generalization of the finite
dimensional case to systems with an infinite number of discrete
oscillators and frequencies. 

In this paper we consider these questions in the context of  Hamiltonian
systems for which the unperturbed dynamics has associated with it 
 discrete {\it and} continuous spectrum.
This situation arises in the study of Hamiltonian PDEs governing 
 functions defined on unbounded spatial domains or, more generally,
  extended systems. 
  The physical picture is that of a system which can
  be viewed as an interaction between one or more discrete oscillators
  and a field or continuous medium.
 In contrast to the KAM theory, where nonresonance implies persistence, we
 find here that resonant nonlinear interaction
 between discrete (bound state) modes and continuum (dispersive
 radiation) modes
  leads to {\it energy transfer} from the discrete to continuum modes.
  This mechanism is responsible for the eventual time-decay and
  nonpersistence of trapped
  states. The rate of time-decay, however, is very slow and hence such a
   trapped state can be thought of as a {\it metastable state}. 
  
  The methods we develop are applicable to  
   a large class of problems which  can be viewed schematically in
  terms of 
  "particle-field interactions". 
  In the present work, 
   we have not attempted to present general results under
   weak hypotheses but rather have endeavored to illustrate, by way
  of example, this widely occurring phenomenon and clearly present the
  strategy for its analysis. A more general point of view will be
   taken up in future work. The approach we use was motivated and in
   part developed  in the context of
   our study of a class of nonlinear Schr\"odinger equations with
   multiple
	nonlinear bound states and the
	quantum resonance problem \cite{kn:SW1bs},
	 \cite{kn:SWseco}, \cite{kn:SWGAFA},  \cite{kn:SW2bs}.  
	  See also \cite{kn:BP1},\cite{kn:BP2} and \cite{kn:PW}.
Related problems are also considered in \cite{kn:LPS} and
\cite{kn:Nier}.

We begin with a linear dispersive Hamiltonian
 PDE for a function $u(x,t)$, $x\in \R^n$ and $t>0$. Suppose that this
system has spatially localized and time-periodic solutions. Such
solutions are often called  {\it bound
states}.  
A typical solution to such a linear system consists of (i) a non-decaying part,
expressible as a linear combination of bound states, plus 
(ii) a part which
decays to zero in suitable norms (dispersion). This paper is devoted
to the study of the following questions:

\medskip
\nit{\bf(1)} Do small amplitude spatially localized and 
 time-periodic solutions persist for
typical nonlinear and Hamiltonian perturbations?

\nit{\bf (2)} What is the character of general small amplitude solutions to
the perturbed dynamics?

\nit{\bf (3)} How are the structures of the unperturbed dynamics manifested
in the perturbed dynamics? 

\medskip

Representative of the class of equations of interest is the nonlinear
Klein-Gordon equation:

\be \D_t^2\ u\ =\ \left(\Delta - V(x) - m^2\right)u + \lambda f(u),\
 \lambda\in\R
\label{eq:nlkg}
\ee
with  $f(u)$ real-valued, smooth in a neighborhood of $u=0$ and 
 having an expansion:
\be f(u) =  u^3 + {\cal O}(u^4) . \label{eq:nonlin}\ee
Here, $u:(x,t)\mapsto u(x,t)\in\R$, $x\in \R^3$, and $t>0$.
We shall restrict our large time asymptotic analysis to the case $f(u)=u^3$. 
 The more general case
(\ref{eq:nonlin}) can be treated by the technique of this paper 
 by making suitable modification of $W^{s,p}$ norms used.

We consider (\ref{eq:nlkg}) with Cauchy data:
\be 
u(x,0)\ =\ u_0(x),\ {\rm and }\ \D_t u(x,0)\ =\ u_1(x).
\nn\ee
Equation (\ref{eq:nlkg}) is a Hamiltonian system with energy:
\be
{\cal E}[u,\D_tu]\ \equiv\ 
{1\over2}\ \int (\D_tu)^2\ +\ |\nabla u|^2\ +\ m^2u^2\ +\ V(x)u^2\
dx\  +\ \lambda \int\ F(u)\ dx, 
\label{eq:energy}
\ee
where $F'(u)=f(u)$ and $F(0)=0$.

In the context of equations of type (\ref{eq:nlkg}), we have found the
following answers to questions (1), (2) and (3).

\nit{\bf (A1)} In a small open neighborhood of the origin, there are no
periodic or quasiperiodic solutions; Corollary 1.1.

\nit{\bf (A2)} All solutions in this neighborhood tend to zero (radiate)
 as $t\to\infty$; Theorem 1.1.

\nit{\bf (A3)} The time decay of solutions is anomalously slow$^*$, {\it i.e.}
a rate which is  slower than the free dispersive rate; Theorem 1.1.
\bigskip

 Dynamical systems of the type we analyze appear
in a number of physical settings.  Consider
a nonlinear medium in which waves can propagate. If the medium has local
inhomogeneities, defects or impurities, these arise in the mathematical
model as a spatially dependent coefficient in the equation 
 ({\it e.g.} localized potential). Such perturbations of the original
homogeneous (translation invariant) dynamics introduce new modes  
 into the system ({\it impurity modes}) which can trap some of the 
 energy and affect
the time evolution of the system; see  
\cite{kn:MS}, \cite{kn:KM}, \cite{kn:ZKMV}. 
\bigskip

Let $\langle K\rangle\ =\ \left( 1\ +\ |K|^2\right)^{1\over2}$.
For the nonlinear Klein-Gordon equation, (\ref{eq:nlkg}), we prove
the following result. 

\begin{theo}
Let $V(x)$ be real-valued and such that
\medskip

\nit {\bf (V1)} For  $\delta>5$ and $|\alpha |\le2$, 
 $|\D^\alpha V(x)|\le C_\alpha\la x\ra^{-\delta}$.

\nit {\bf (V2)} 
 $(-\Delta +1)^{-1}\left( (x\cdot\nabla)^l V(x)\right)(-\Delta +1)^{-1}$ is
bounded on $L^2$ for $|l|\le N_*$ with $N_*\ge10$.

\nit {\bf (V3)}  Zero is not a {\it resonance} of the operator $-\Delta +V$; see
\cite{kn:JK}, \cite{kn:Y}.

Assume the operator 
\be B^2= - \Delta + V(x) + m^2 \label{eq:Bdef}\ee
has continuous spectrum, $\sigma_{cont}(H)= [m^2,\infty)$, and 
 a unique strictly positive simple eigenvalue, $\Omega^2<m^2$ with 
associated normalized eigenfunction, $\varphi$:
\be B^2\varphi = \Omega^2\varphi.\nonumber\ee
Correspondingly,
 the linear Klein-Gordon equation (\ref{eq:nlkg}), with $\lambda=0$, has
a two-parameter family of spatially localized and time-periodic
solutions of the form: 
\be u(x,t)\ = \ R\ \cos(\Omega t + \theta)\ \varphi(x).\label{eq:exact}\ee
Assume the resonance condition 
\be 
\Gamma\equiv 
{\pi\over3\Omega}\ \left(\Pc\varphi^3,\delta(B-3\Omega)\Pc\varphi^3\right)
\equiv\ {\pi\over3\Omega}\ \left|\left({\cal F}_c\varphi^3\right)(3\Omega)\right|^2
 > 0.\label{eq:nlfgr}\ee
Here, $\Pc$  denotes  the projection onto the continuous spectral 
 part of $B$ and ${\cal F}_c$ denotes  the  Fourier transform relative to the 
continuous spectral part of $B$.

 Assume that the initial data $u_0$, $u_1$ are such that
 the norms  $\|u_0\|_{W^{2,2}\cap W^{2,1}}$ and 
  $\|u_1\|_{W^{1,2}\cap W^{1,1}}$  are sufficiently
small. 
Then, the solution of the
initial value problem for (\ref{eq:nlkg}),
 with $\lambda\ne0$ and $f(u)=u^3$  decays as $t\to\pm\infty$. 
The solution $u(x,t)$ has the following expansion as $t\to\pm\infty$:
\be u(x,t)\ =\  
R(t)\ \cos\left(\Omega t +
\theta (t) 
 \right)\ \varphi(x)\ +\ \eta(x,t), \nonumber 
\ee
where
\be
R(t) =  {\cal O}(|t|^{-{1\over4}})\ , 
\theta (t) =  {\cal O}(|t|^{1\over2}),\ {\rm and}\ 
\| \eta(\cdot ,t) \|_8 = {\cal O}(|t|^{-{3\over4}}).
 \label{eq:uasymp}
\ee
 More precisely,
 \ba
 R\ &=&\ \tilde R\ +\ {\cal O}(|\tilde R|^2),\ ( 
 \ |\tilde R| \ {\rm small})
 \nn\\
 \tilde R\ &=&\
  2^{1\over4}\tilde R_0\ 
   ({1+{3\over4}\tilde R_0^4\ \Omega^{-1}\ \lambda^2\ \Gamma\ |t|})^{-{1\over4}}
 \cdot\left( 1+{\cal O}(|t|^{-\delta})\right),\ \delta >0\nn\\
 R(0)\ &=&\ R_0,\ R_0^2\ =\ \left|\left(\varphi,u_0\right)\right|^2
						   \ +\ \Omega^{-2}\left|\left(\varphi,
						   u_1\right)\right|^2\nn
\ea
\end{theo}

\begin{cor}
Under the hypotheses of Theorem 1.1, 
there are no periodic or quasiperiodic orbits of the flow $t\mapsto
\left( u(t),\D_tu(t)\right)$ defined by (\ref{eq:nlkg}) in a 
 sufficiently small neighborhood of the origin in the space 
 $\left( W^{2,1}\cap W^{2,2}\right) \times \left( W^{1,2}\cap W^{1,1}\right)$. 
\end{cor}

It is  interesting to contrast our results with those known for
Hamiltonian  partial differential equations for a function $u(x,t)$,
 where $x$ varies over a
 compact spatial domain, {\it e.g.} periodic or Dirichlet boundary
 conditions \cite{kn:Bgn}, \cite{kn:CW}, \cite{kn:Kuk}.
  For nonlinear wave equations of the
   form,
   (\ref{eq:nlkg}), with {\it periodic boundary conditions} in $x$,
	KAM type results  have been proved; invariant tori, associated with
	a
	 {\it nonresonance} condition persist under
	  small perturbations.
	   The {\it nonresonance}
		 hypotheses of such results fail in the current context
		 , a consequence of
		 the  continuous spectrum associated with unbounded spatial
		 domains.
		  In our situation, non-vanishing {\it resonant coupling} (condition
		  (\ref{eq:nlfgr}))
			provides the mechanism for the radiative decay  and
			therefore
			   nonpersistence of
				localized periodic solutions.

\noindent 
{\bf Remarks:} 

\noindent{(1)}    
 The condition (\ref{eq:nlfgr}) is a {\it nonlinear variant of
 the Fermi golden rule} arising in quantum mechanics; see, for example,
 \cite{kn:CFKS},
 \cite{kn:SWGAFA}.
This condition holds generically in the space of potentials satisfying
the hypothesis of the theorem.
 Note that the condition  (\ref{eq:nlfgr}) implies that 
\be 3\Omega\in  \sigma_{cont}(B) \label{eq:hrc}\ee 
 {\it i.e.} the frequency, $3\Omega$,
 generated by the cubic nonlinearity lies in the continuous spectrum of
 $B$.
The hypothesis (\ref{eq:nlfgr}) ensures a nonvanishing coupling to the
continuous spectrum.

\nit {(2)} The regularity and decay hypotheses on $V(x)$ are related to the
techniques we use to obtain suitable decay estimates on the linear evolution in 
$L^2(\la x\ra^{-\sigma} dx)$ and $L^p$; see section 2.

\nit (3) {\it Persistence of dynamically
 stable bound states for special perturbations:}
  There are nonlinear perturbations of the linear Klein-Gordon equation,
   (\ref{eq:nlkg}) with $\lambda =0$, for which
	there is a  persistence of time-periodic and spatially
	localized
	 solutions. Suppose we extend our considerations to the class of
	 complex-valued
	  solutions: $u:\R^n\times\R\to \C$, and  study the perturbed
	  equation:
	   \be
	\left(\ \D_t^2\ -\ \Delta\ +\ V(x)\ +\ \lambda |u|^{p-1}\
	\right)u\ =\ 0,\ \label{eq:complex-nlkg}
   \ee
where $1<p<\infty$ for $n=1,2$ and $1<p< {{n+2}\over{n-2}}$ for
$n\ge3$. Unlike (\ref{eq:nlkg}), equation (\ref{eq:complex-nlkg})
 has the symmetry $u\mapsto ue^{i\gamma}$ and it has been shown \cite{kn:RW} 
  that
  (\ref{eq:complex-nlkg})
  has time periodic and spatially localized solutions of
   the form
$e^{i\omega t}\Phi(x;\omega)$ with $\Phi\in H^1$, which
bifurcate
 from the zero solution at the  point eigenvalue of
 $-\Delta + V(x) -\omega^2$, by global variational
 \cite{kn:PLL}, \cite{kn:strauss} and local bifurcation methods  
 \cite{kn:Nirenberg}. There are numerous other examples of equations for
  which the persistence of coherent structures under perturbations is
  linked to the perturbation respecting a certain symmetry of the
  unperturbed problem.
	The stability of small amplitude bifurcating
	 states of (\ref{eq:complex-nlkg}) can be proved using the methods of 
	 \cite{kn:GSS}, \cite{kn:MIW86}. An asymptotic
	  stability and 
	scattering theory has been developed for  
	  nonlinear Schr\"odinger dynamics (NLS) in
	 \cite{kn:SW1bs}. If the
	 potential $V(x)$ supports more than one bound
	state, the above methods 
	  can be used to show the persistence of
	{\it nonlinear excited states}. However, it is
	 shown in \cite{kn:SW2bs} that
	the NLS excited states are unstable, due to a
	resonant mechanism of the
	type studied here for (\ref{eq:nlkg}).

\noindent{(4)}  
 In contrast with the time-decay rates associated with linear dispersive
 equations,   the time-decay rates  
  of typical solutions described by Theorem 1.1 is anomalously slow. See
  section 8 for a discussion of this point.
\bigskip

We now present an outline of the ideas behind our analysis.
For $\lambda=0$, solutions of equation (\ref{eq:nlkg}) are naturally
decomposed into their discrete and continuous spectral components:
\be u(x,t)\ =\ a(t)\varphi(x)\ +\ \eta(x,t),
\quad \left(\varphi,\eta(\cdot,t)\right)\ =\ 0, \label{eq:decomp}\ee
where $(f,g)$ denotes the usual complex inner product on $L^2$.  
The functions $a(t)$ and 
$\eta(x,t)$ satisfy system of {\it decoupled} equations:
\ba 
 a'' + \Omega^2\ a &=& 0, \\
\D_t^2\ \eta\ +\ B^2\eta &=& 0,
\ea
with initial data
\ba
a(0)\ &=&\ \left(\varphi,u_0\right),\ \quad a'(0)\ =\
\left(\varphi,u_1\right)\nn\\
\eta(x,0)\ &=&\ \Pc u_0,\ \quad  \D_t\ \eta(x,0)\ =\ \Pc u_1
\label{eq:aetadata}
\ea 

For $\lambda\ne0$, and for small amplitude initial conditions, we use
the same decomposition, (\ref{eq:decomp}). Now the discrete and the
continuum modes are coupled and the dynamics are qualitatively described
by the following {\it model system}:
\ba 
a'' + \Omega^2\ a &=& 
 3\lambda a^2\left(\chi,\eta(\cdot ,t)\right) \label{eq:modelcce1} \\ 
\D_t^2\ \eta -\ \Delta \eta + m^2\eta &=& \lambda\ a^3\chi
\label{eq:modelcce2}.
\ea
Here, $\chi(x)$ is a localized function of $x$; in particular,
$\chi={\varphi^{3}}$, and we assume (see (\ref{eq:nlfgr})) $3\Omega > m$.
In selecting the model problem (\ref{eq:modelcce1}-\ref{eq:modelcce2}),
we have replaced $B^2$, restricted to its continuous spectral part,  by 
$B_0^2=-\Delta + m^2$, which intuitively should lead to the same
qualtitative result.

The system (\ref{eq:modelcce1}-\ref{eq:modelcce2}) 
 can be interpreted as a system   
governing the dynamics of discrete oscillator, with amplitude $a(t)$
and natural frequency $\Omega$,  coupled
to a continuous medium in which waves, of amplitude $\eta(x,t)$, propagate, 
or as an oscillating particle coupled to a field.
\footnote{A related example is a model introduced in 1900 by H. Lamb
\cite{kn:Lamb}, governing the
oscillations of mass-spring-string system. See also  
\cite{kn:BK}, \cite{kn:FKM}, \cite{kn:SDG}, \cite{kn:JP}.}  

The system (\ref{eq:modelcce1}-\ref{eq:modelcce2}) 
 is a Hamiltonian system with conserved
total energy functional:
\ba \tilde{\cal E}[\ \eta,\D_t\eta, a,a'\ ] &\equiv& 
 \frac{1}{2}\int\ (\D_t\eta)^2+|\nabla\eta |^2\ +\ m^2 \eta^2\ dx\ +\
\frac{1}{2}\left({a'}^2 + \Omega^2 a^2 \right)
 \nonumber \\
&-&\lambda a^3\int\chi(x)\eta(x,t)\ dx \label{eq:conserve}\ea

We now solve  (\ref{eq:modelcce2}) and substitute the result into
(\ref{eq:modelcce1}). 
Note that, to leading order, solutions of  (\ref{eq:modelcce1})
oscillate with frequency $\Omega$. Therefore the $a^3$ term
in (\ref{eq:modelcce2}) acts as an external driving force with a $3\Omega$
frequency component. 
 Since $9\Omega^2$\  is larger than $m^2$, a nonlinear
resonance of the oscillator with the continuum takes place. 
 To calculate the effect of this resonance requires a careful analysis
involving  (i) a study of 
 singular limits of
resolvents as an eigenvalue parameter approaches the continuous
spectrum (see section 4) and (ii) and the derivation of a normal form which is
 natural for an infinite dimensional conservative system with
dispersion   
(see section 5).  
This leads to an equation of the following type for $a(t)$ (or rather some
near-identity transform of it):
\be a'' + \left(\Omega^2 + {\cal O}(\vert a\vert^2)  
 \right)a = 
-\Gamma\ r^4\ a',\ \ t>0\label{eq:dosc}\ee
where $ r = {\cal O}(|a|)$, and $\Gamma>0$ is the positive number
given in (\ref{eq:nlfgr}). For the model system (\ref{eq:modelcce2}),
${\cal F}_c$ in (\ref{eq:nlfgr}) is replaced by the usual Fourier
transform. 
This is the equation of a nonlinearly damped oscillator:
\be {d\over dt}\left((a')^2 + [\Omega^2 + {\cal O}(|a|^2)] a^2\right) =
 -2\Gamma r^4 (a')^2\ <0,\ {\rm for\ } \ t>0\nonumber\ee
Therefore, nonlinear resonance is responsible for {\it internal damping}
in the system; energy is lost or rather transferred 
 from the discrete oscillator into the field or 
continuous medium, where it is propagated to infinity as dispersive
waves. Solutions of
(\ref{eq:dosc}) decay with a rate ${\cal O}(t^{-1/4})$, and it follows
that $\|\eta(\cdot ,t)\|_\infty = {\cal O}(t^{-3/4}).$ Note, however
 that the
total energy of the system, an $H^1$ type quantity, is conserved. 
In (\ref{eq:dosc}), we have neglected 
higher order effects coming from the continuous spectral part of $H$. These, it
turns out, has a small effect and can be treated perturbatively. 

Aspects of the analysis are related to our recent treatment of the
quantum resonance problem \cite{kn:SWseco}, \cite{kn:SWGAFA}. 
  The situation with quantum
resonances can be summarized briefly as follows. Let $H_0$ be   
a self-adjoint operator having an eigenvalue, $\lambda_0$, embedded in
its continuous spectrum, with corresponding normalized eigenfunction $\psi_0$.
 Let $W$ be a small localized symmetric perturbation, 
 satisfying the (generically valid) Fermi golden rule resonance condition:
\be \Gamma_0\ \equiv\ \pi\left(W\psi_0,\delta(H_0-\lambda_0) \Pc W\psi_0\right)\ne0.
\label{eq:fgr}\ee
Then in a neighborhood of $\lambda_0$, we show that the spectrum of 
 $H=H_0+W$ is
absolutely continuous by proving that all solutions of the Schr\"odinger
equation:
\be
i\ \D_t \psi\ =\ H \psi\ =\ (H_0 + W) \psi, \label{eq:ls}
\ee
 with initial data which is spectrally localized (with
respect to $H$) in a neighborhood of 
$\lambda_0$, decay to zero in a local energy norm as $t\to\pm\infty$.

Such solutions are characterized by exponential time-decay for an
initial transient period, and then algebraic (dispersive decay)
thereafter. During this transient period, $A(t)$, the projection onto the mode
 $\psi_0$, is governed by the equation:
\be {A'}(t) = (i\lambda_* - \Gamma_0)A(t),\ \ t>0 \label{eq:linA}\ee
where $\lambda_* \sim \lambda_0 + \left(\psi_0,W\psi_0\right)$, and 
by (\ref{eq:fgr}), $\Gamma_0>0$.

In the class of nonlinear problems under consideration, if we express
the amplitude, $a(t)$, as 
\be 
a(t)\ =\ A(t)\ e^{i\Omega t} + \ov (t)\ e^{-i\Omega t}
\label{eq:aA}
\ee
then we find after a near-identity transformation an equation of the
form:
\be A' = ic_{21}|A|^2A\ +\ (ic_{32} - {3\over4}{\lambda^2\over\Omega}\Gamma)|A|^4A,\quad t>0.\label{eq:Aosc}\ee
From this, the ${\cal O}(t^{-1/4})$ behavior is evident. 
Equations (\ref{eq:aA}-\ref{eq:Aosc}) lead to  (\ref{eq:dosc}).  
As in (\ref{eq:dosc}), we have in (\ref{eq:linA}) and (\ref{eq:Aosc})
neglected the higher order
 coupling to the continuous spectral (radiation) component
of the solution. These contributions are treated perturbatively.

\nit {\bf Remarks:}    

\nit{(1)} {\it Lamb shift}:\  
Note that in the nonlinear problem, asymptotically there is no {\it Lamb
shift} type correction to the frequency; the frequency shift is ${\cal
O}(|A|^2)={\cal O}(t^{-1/2})$ as $t\to\pm\infty$.

\nit{(2)}
{\it Emergence of irreversible behavior from reversible dynamics;
 dissipation through dispersion}:\ 
Being Hamiltonian, the underlying equation of motion, 
 (\ref{eq:nlkg}), is time reversible. In particular, the equation has
the invariance: $u(x,t)\mapsto u(x,-t)$.  Yet, the equation in  
(\ref{eq:Aosc}) is clearly not time-reversible. This apparent paradox
is related to the "$\varepsilon$-{\it prescription}" discussed in section
4 and  Proposition 2.1;  the singular limit
\be \lim_{\veps\downarrow 0}\ \exp(i\sqrt{-\Delta + m^2}\ t)(-\Delta +m^2 -
 E \pm i \veps)^{-1}\ee
satisfies a local decay estimate as $t\to\pm\infty$. For $t<0$, the
corresponding equation for $A(t)$ would have $-\Gamma$ replaced by  
$+\Gamma$; {\it cf.} \cite{kn:AB}, \cite{kn:BP2}, \cite{kn:SWseco},
\cite{kn:SWGAFA}.
\bigskip

%- $\phi^4$ Breathers: Daschen et al breather to all orders; see also (?)
% a paper by DK Campbell
%-  Kruskal-Segur beyond all orders defect 
%
% These can be treated by our methods, 
%-  other examples: NLS with 3rd order dispersive effects (Wai et al), 
% KdV with 5th
% order surface tension effect (Hunter-Scheurle), 
%  Euler equations with surface tension (Beale, Amick(?)-Toland)
%
\bigskip

The nonpersistence of   
 small amplitude spatially localized and   time-periodic 
 or quasiperiodic
solutions to (\ref{eq:nlkg}) has been studied in 
\cite{kn:sigalcmp},\cite{kn:sigjapan}. In this work, the phenomenon of
nonpersistence is  
formulated as a question concerning the instability of embedded eigenvalues of a
suitable {\it linear} self-adjoint operator. 
A result on structural instability is proved; 
  under the hypothesis (\ref{eq:nlfgr}),  
 time-periodic or quasiperiodic solutions  of the linear problem equation
($\lambda = 0$) do not continue to nearby solutions of the nonlinear
problem ($\lambda\ne0$). 
 
The question of nonpersistence of periodic solutions has 
  also been considered extensively in the context of 
the sine-Gordon equation
\be \D_t^2 u = \D_x^2 u - \sin u. \label{eq:sg}\ee
Spatially localized and time-periodic
solutions of the sine-Gordon equation are called {\it breathers} and the
 question of their persistence under small Hamiltonian perturbations in the
dynamics  has been the subject of extensive investigations. See, for
example, \cite{kn:SK},
 \cite{kn:C}, \cite{kn:CP}
  \cite{kn:WeA}, 
\cite{kn:BMW}, \cite{kn:Bnr},
 \cite{kn:D},  
  \cite{kn:Kich}, 
 \cite{kn:PS}, \cite{kn:St}, \cite{kn:McSh}.

  Analytical, formal asymptotic and numerical studies 
  strongly suggest that for  
typical Hamiltonian perturbations of the sine-Gordon equation, for
example, the $\phi^4$ model: 
\be \D_t^2\ u = \D_x^2 u -  u + u^3, \label{eq:phi4}\ee
 no small amplitude breathers exist and that solutions obtained from
 spatially localized initial data {\it radiate} 
 to zero
very slowly as $t$ tends to infinity.
\footnote{ In \cite{kn:MA}, breather type solutions have been 
 constructed for the discrete
sine-Gordon equation, where $\D_x^2$, is replace by its discretization
 on a sufficiently coarse lattice. Also, a generalization of 
 the notion of breather  
has been considered in the  geometric context of {\it
wave maps} \cite{kn:SS}.} 
We believe this is related to the mechanism for slow radiative decay, as  
explained by Theorem 1.1 in the context of (\ref{eq:nlkg}).

%%%%%%%%%%%%%%%%%%%%%%%%
Finally, we wish to comment on the connection between our work and 
 the approach taken
in  \cite{kn:sigalcmp}, \cite{kn:BMW}, \cite{kn:D}, and 
 \cite{kn:Kich}.  
 As in the continuation theory of periodic solutions of ordinary
differential equations \cite{kn:CL}, it is natural to 
 seek a periodic solution of (\ref{eq:nlkg}) for 
$\lambda\ne0$ which behaves, as the amplitude $a$ tends to zero, like
 a solution (\ref{eq:exact}) of the linear limit problem. 
 The equation is {\it autonomous} with respect to time, so we seek a
 $2\pi\Omega_a^{-1}$-periodic solution for $\lambda\ne0$ with an
amplitude dependent period. Since we do not know the period {\it \'a priori},
 it is convenient to define 
\be
u(x,t)\ =\ U_a(x,s), \quad\ s=\Omega_a t
\nn\ee
and require that $U_a(x,s)$ be  $2\pi$ periodic in $s$. Thus
(\ref{eq:nlkg}) becomes
\be
\left( \Omega_a^2\ \D_s^2\ +\ B^2\ \right)\ U_a\ =\ \lambda\ U_a^3.
\label{eq:s-nlkg}
\ee
We formally expand the solution and frequency:
\ba
U_a\ &=&\ aU_1\  +\ a^3\ U_3\ +\ ...\nn\\
\Omega_a &=&\ \Omega\  +\ a^2\ \Omega_2\ +\ ...\ .
\label{eq:Ua}\ea
%where $U_3=U_3^{(1)}+U_3^{(2)}$,
Substitution of (\ref{eq:Ua}) into (\ref{eq:s-nlkg}) and 
assembling terms according to their order in $a$, one gets a hierarchy of 
 equations beginning with
\ba
{\cal O}(a^1):\qquad 
 \left( \Omega^2\D_s^2\ +\ B^2\ \right)\ U_1\ &=&\ 0\label{eq:U1}\\
{\cal O}(a^3):\qquad  
\left( \Omega^2\D_s^2\ +\ B^2\ \right)\ U_3\ &=&\ 
 \lambda U_1^3\ -\ 2\Omega\Omega_2\D_s^2U_1
 \label{eq:U3}
 \ea
Equation (\ref{eq:U1}) has a solution
\be U_1(x,s)\ =\  \cos{s}\ \varphi(x).\nn\ee
Substitution into  (\ref{eq:U3}) gives the following explicit equation
for $U_3$:
 \be
 \left( \Omega^2\D_s^2\ +\ B^2\ \right)\ U_3\ =\ 
     \left(  {3\lambda\over4}\varphi^3\ +\ 2\Omega\Omega_2\varphi\ \right)\ 
           \cos{s}\ + {\lambda\over4}\varphi^3\cos{3s} 
\label{eq:U3a}
\ee
We now express $U_3$ in the form $U_3=U_3^{(1)} + U_3^{(2)}$, where
\ba
\left( \Omega^2\D_s^2\ +\ B^2\ \right)\ U_3^{(1)}\ &=&\ 
 \left(  {3\lambda\over4}\varphi^3\ +\ 2\Omega\Omega_2\varphi\ \right)\
			\cos{s}\label{eq:U3i}\\
 \left( \Omega^2\D_s^2\ +\ B^2\ \right)\ U_3^{(2)}\ &=&\
 {\lambda\over4}\varphi^3\cos{3s}.\label{eq:U3ii}
 \ea
Since the inhomogeneous term in (\ref{eq:U3i}) is nonresonant with the
 continuous spectrum of $B^2$, one can find a $2\pi$-periodic  solution
$U_3^{(1)}$ provided the right hand side is  
 $L^2(S^1_{2\pi}\times\R^3)$ - orthogonal to the adjoint zero mode: 
$\cos{s}\ \varphi(x)$. This latter condition uniquely determines $\Omega_2$. 
 Note however that the right hand side of
(\ref{eq:U3ii}) is resonant with the continuous spectrum of $B^2$ if
$3\Omega > m$. To compute the obstruction to solvability, seek
a solution of (\ref{eq:U3ii}) of the form
$U_3^{(2)}\ =\ \cos{3s}\ F$. Then,
\be
\left(\ B^2\ -\ 9\Omega^2\ \right)F\ =\ {\lambda\over4}\varphi^3\nn
\ee
and thus we expect to find $F\in L^2$ if and only if the component of
$\varphi^3$ in the direction of the generalized eigenfunction at
frequency $3\Omega$ of the operator $B$ vanishes. Therefore, if the
nonlinear Fermi golden rule (\ref{eq:nlfgr}) holds, our formal expansion
in $L^2(S^1_{2\pi}\times\R^3)$ breaks down. 
 The relation with our work is that we 
prove that this obstruction to solvability, in fact, implies 
 the radiative behavior of solutions
described in our main theorem. 

\bigskip
\noindent{\bf Acknowledgments:} 
The problem studied in this paper was raised by T.C. Spencer in
lectures and informal discussions in the late 1980's. 
M.I.W. learned of this problem from T.C. Spencer  and A.S. from I.M. Sigal. 
The authors wish to thank them for their insights and continued
interest. 
 The authors also wish
to thank B. Birnir for stimulating 
discussions, and  R. Pyke and J.B. Rauch for their careful reading
 of and thoughtful comments on the manuscript. 
 This research was carried out while MIW was on sabbatical
leave in the Program in Applied and Computational Mathematics at 
 Princeton University. MIW would like to thank Phil Holmes for his
hospitality and for providing a stimulating research environment. 
This research was supported in part by  NSF grant DMS-9401777 and an 
FAS-Rutgers grant award (AS) and by NSF grant 
 DMS-9500997 (MIW).

\bigskip
\section{Linear Analysis}
\bigskip

In this section we summarize the tools of linear analysis
  employed in this paper. 

\subsection{ Estimates for the (free) Klein Gordon equation}

 We begin by considering the Cauchy problem for the linear Klein-Gordon 
equation for a function $u(x,t),\ x\in \R^n,\ t\ne0$.
\ba
\partial_t^2 u\ &=&\ (\Delta-m^2)u=-B_0^2 u\label{eq:lkg}\\
 u(x,0)\ &=&\ u_0(x),\ \  \partial_t u(x,0)=u_1(x)\ .\label{eq:lkgdata}
\ea
There exist operators $\ezz$ and $\eoz$ such that
\be
u(x,t)\ =\ \ezz\ u_0\  +\ \eoz\ u_1.  \label{eq:lkgsoln}
\ee
We write
\ba \ezz f &=&\ \cos B_0t\  f\ , \rm{and} \nn\\
\eoz g\ &=&\ {\sin B_0t\over B_0} g\ ,\label{eq:3.4}
\ea
where ($\omega(k)\ =\ \sqrt{m^2+k^2}$):
\ba
\ezz f\ &=&\ \int\ \cos\omega(k)t\ \hat f(k)\ e^{ik\cdot x}\ dk\nn\\
\eoz g\ &=&\ \int\ {\sin\omega(k)t\over\omega(k)}\ \hat g(k)\ e^{ik\cdot
x}\ dk.
\label{eq:ezeo}
\ea
The first result we cite is proved using stationary phase methods; see 
\cite{kn:Br}. See \cite{kn:Kl} for another approach to decay estimates
for (\ref{eq:lkg}).
 Let $W^{s,p}(\R^n)$ denote the Sobolev space of functions with
 derivatives of order $\le s$ in $L^p$. 

\begin{theo}
Let $1<p\le 2$,\  ${1\over p}+{1\over p'}=1$,\  $\delta\equiv
{1\over2}-{1\over p'}$ and $0\le\theta\le1$. Then, 

\nit (a)\ if $\delta(n+1+\theta)\le \nu +s$, we have for $\nu=0,1$:
\ba
\|\ E_\nu^{(0)} g\|_{p'}\ &\le&\ K(t)\ \|g\|_{s,p}\ ,\ t\ge0,\ 
 {\rm where } \label{eq:eozest}
 \\ 
 K(t)\ &=&\ C t^{-(n-1-\theta)\delta},\ 0<t\le1,\nn\\
      &=&\ C t^{-(n-1+\theta)\delta}, \ t\ge1. 
\label{eq:ezzest}
\ea

\nit (b) 
If $s\ge (1/2-1/p')(n+2)-1$, then for $t\ge1$ the (Schr\"odinger-like) $L^{p'}$
decay rate,  
  $t^{-n({1\over2}-{1\over p'})}$, holds in  (\ref{eq:ezzest}).
\end{theo}

In subsequent sections, we shall use some specific consequences of this
result. 

\begin{cor} 
Consider the linear Klein Gordon equation (\ref{eq:lkg}) in dimension
$n=3$. Then, 

\ba 
||\eoz g||_8 &\le& C\ t^{-{9\over 8}}\ ||g||_{1,{8\over7}}\ 
 \ t\ge 1 \label{eq:eoztge1}\\
||\eoz g||_8\ &\le&\  C\ t^{-{3\over 8}}\ ||g||_{1,{8\over7}}\ 
 0<t\le 1
  \label{eq:eoztle1}\\
|| E^{(0)}_\nu (t) g||_4\ &\le&\  C\ t^{-{1\over 2}}\  ||g||_{1,{4\over3}},\ 
 \ t > 0,\  
 \nu\ =\ 0,1.
       \label{eq:ezjtge1}
\ea
\end{cor}

\nit {\it proof:} Estimates (\ref{eq:eoztge1}) and (\ref{eq:eoztle1})
follow from the theorem  
 with the choice of parameters: $p'=8$,\ $s=1$,\ $n=3$, and $\theta
=1$. Estimate (\ref{eq:ezjtge1}) follows  from the theorem with the choice of
parameters: 
\ $p'=4$,\ $s=1$, $n=3$\ and $\theta=0$.
\bigskip

\subsection{Estimates for the Klein Gordon equation with a
potential} 
\medskip

We now consider the Cauchy problem for the linear 
  Klein-Gordon equation with a potential
\be
\partial^2_t u\ =\ (\Delta-m^2-V(x))u\ =\ -B^2u
 \label{eq:lkgv}
\ee
where $B^2$ is positive and self-adjoint. 

We write the solution to the initial value problem for (\ref{eq:lkgv})
with initial data (\ref{eq:lkgdata}) as:
\ba
u(x,t)\ =\ \ez u_0 &+& \eo u_1\ ,\ {\rm where} \\
\smallskip
\ez f\ &=&\ \cos(Bt)f\ ,\ \rm{and}\\
\eo g\ &=&\ {\sin (Bt)\over B} g\ .\nn
\ea

One expects that estimates of the type appearing in Theorem 2.1 and
Corollary 2.1 will hold as well for $\ez$ and $\eo$ restricted to $\Pc
L^2$, the  
continuous spectral part of $H = B^2$. One can relate functions of the
operator $H$, on its continuous spectral part, to functions of $H_0\ =\
B_0^2$
using {\it wave operators}. Let
\be
W_+\ =\ {\rm strong}\ -\ \lim_{t\to\infty}\ e^{itH}e^{-itH_0}.
\label{eq:waveop}
\ee  
Wave operators relate $H_0$ to $H$ on $\Pc L^2$ via the intertwining property:
\be
H\  = W_+\ H_0\ W_+^*\quad {\rm on }\ \Pc L^2.
\label{eq:intertwine}
\ee
 K. Yajima \cite{kn:Y} has proved the $W^{k,p}(\R^n)$ 
 boundedness of wave operators for the
Schr\"odinger equation. A consequence of this work is the following
result for spatial dimension $n=3$:

\begin{theo} 
  Let $H=-\Delta+V$, 
where $V(x)$ be real-valued function on $\R^3$
satisfying the following hypotheses, which are satisfied by smooth and 
 sufficiently rapidly decaying
potentials:

\nit For any $|\alpha|\le\ell$ there is a constant $C_\alpha$ such that
$$\bigg|{\D^\alpha V\over\partial x^\alpha}(x)\bigg|\ \le\ C_\alpha\langle x\rangle^{-\delta}\
,\quad \delta> 5\ .$$
Additionally, assume that $0$ is neither an eigenvalue nor a  {\it resonance} of
$H$; see \cite{kn:Y}, \cite{kn:JK}.
If $0\le k,k'\le l$ and $1\le p,p'\le\infty$, then there exists a constant $C>0$ such
that for any Borel function $f$  on ${\bf R}^1$ we have
$$C^{-1}\ ||f(H_0)||_{B(W^{k,p}, W^{k',p'})}\ 
 \le\ ||f(H)\Pc(H)||_{B(W^{k,p}, W^{k',p'})}\ \le\ 
 C\ ||f(H_0)||_{B(W^{k,p}, W^{k',p'})}.$$
Here,
$\Pc(H)$ denotes the projection onto the continuous spectral part of 
 the operator $H$.
\end{theo}

\nit {\bf Remark:}  In our applications, we shall use Theorem 2.2 with
 $|l|\le2$.

Theorem 2.2 implies that the dispersive estimates for the free
Klein Gordon group carry over to the operators $\ez$ and $\eo$ restricted
to the continuous spectral part of $B$.

\begin{theo}
 If $g\in \ \hbox{\rm Range}\ \Pc(H)$, then each of the estimates of Theorem 2.1 and Corollary 2.1 hold with
$E^{(0)}_j$ replaced by $E_j$.
\end{theo}

\bigskip
\subsection{ Singular resolvents and time decay}

As discussed in the introduction, the damping term in the effective nonlinear
oscillator (\ref{eq:Aosc}), is related to the evaluation of a singular limit
of the resolvent as the eigenvalue parameter approaches the continuous
spectrum. To ensure that the correction terms to the nonlinear
oscillator (\ref{eq:Aosc}) can be treated perturbatively, we require
  local decay estimates for the operator 
   $e^{iBt}(B-\Lambda\pm i0)^{-1}$, where
 $\Lambda$ is a point in the interior of the continuous spectrum
  of $B$ ($\Lambda >m$).
 Such estimates are analogous to those used by the authors in recent
 work on a time dependent approach to the quantum resonance problem
\cite{kn:SWGAFA}.

We begin with a proposition, which is essentially a restatement of
Theorem 2.3 for $E_j(t)$, and is proved the same way.

\begin{prop} ({\it $W^{s,p}$ estimates for $e^{iBt}$}) 
 Assume that $V(x)$ satisfies the hypotheses of Theorem 2.2 ($n=3$);
 see \cite{kn:Y} for hypotheses for general space dimension, $n\ge3$.
 Let $p$ and $p'$ be as in Theorem 2.1 and let $l\ge s\ge
 ({1\over2}-{1\over p'})(n+2)$, where $l$ is as in Theorem 2.2. Then,
\be
  \|\ e^{iBt}\ 
    \Pc\ \psi\|_{p'}
 \ \le\
   C|t|^{-n({1\over2}-{1\over p'})}\ \|\psi\|_{s,p},\ t\ne0,
   \label{eq:Wsp}
   \ee
   \end{prop}

The next proposition is the "singular" resolvent estimate we shall
require in section 7. Let $\sigma_*(n)=\max\{{n\over2},2+{2n\over
n+2}\}$.
\begin{prop} Assume the hypotheses of Proposition 2.1. Additionally, assume
that $V(x)$ satisfies hypothesis {\bf (V2)} of Theorem 1.1.  Also, let  
  $\sigma>\sigma_*(n)$.
 Then, for any point $\Lambda$ ($\Lambda
> m$) in the continuous spectrum of $B$ we have for $l=0,1,2$:
\ba
\|\la x\ra^{-\sigma}\ e^{iBt}\ \left(B-\Lambda +
 i0\right)^{-l}\ \Pc\ \la x\ra^{-\sigma}\ \psi\|_{2}
 \ &\le&\ 
C\ \la t\ra^{-{2n\over n+2}}\ \|\psi\|_{1,2},\ t>0,\nn\\
\|\ \la x\ra^{-\sigma}\ e^{iBt}\ \left(B-\Lambda -
 i0\right)^{-l}\ \Pc\ \la x\ra^{-\sigma} \psi\|_{2}
  \ &\le&\
  C\ \la t\ra^{-{2n\over n+2}}\ \|\psi\|_{1,2},\ t<0,
\label{eq:ld2}
\ea
\end{prop}

\nit{\it proof}: We prove the estimate (\ref{eq:ld2}) for the case of
the $t>0$.  For the case, $t<0$,  
 the same argument with simple modifications applies.

 Let $g_\Delta = g_\Delta(B)$  
 denote a smooth  
characteristic functions of an open interval, $\Delta$
  which contains 
$\Lambda$ and is contained in the continuous spectrum of $B$. 
  Also, let 
  ${\overline g}_\Delta\equiv 1 - g_\Delta$. We then
write, for $l=1,2$:
\ba
e^{iBt}\ \left( B - \Lambda + i\veps\right)^{-l}\Pc\ &=&\ 
e^{iBt}\ g_\Delta(B)\ \left( B - \Lambda + i\veps\right)^{-l}\Pc\
+\  e^{iBt}\ \overline g_\Delta(B)\ \left( B - \Lambda +
 i\veps\right)^{-l}\Pc\nn \\
&\equiv&\ S_1^\veps(t) \ +\ S_2^\veps(t).\label{eq:s1s2}
\ea

We now estimate the operators $S_j\equiv \lim_{\veps\to0}S_j^\veps$,
$j=1,2$ individually. 
\bigskip

\centerline{\it Estimation of $S_1(t)$:}
\medskip

Consider the case $l=2$; the case $l=1$ is simpler.
First, we note that:
\ba
&&\la x\ra^{-\sigma}\ S_1^\veps(t)\ \la x\ra^{-\sigma}\ \nn\\
\ \ \ &=&\ -e^{it(\Lambda -i\veps)}\ \int_t^\infty \ ds\ \int_s^\infty\
 d\tau\  
  \la x\ra^{-\sigma}\  e^{i\tau(B-\Lambda +i\veps)}\
 g_\Delta(B)\Pc\ \la x\ra^{-\sigma}\ \label{eq:integralrep}
 \ea
 Now we claim that the operator in the integrand satisfies the estimate: 
 \be
 \| \la x\ra^{-\sigma}\  e^{i\tau(B-\Lambda +i\veps)}\
  g_\Delta(B)\Pc\ \la x\ra^{-\sigma}\|_{{\cal B}(L^2)}\ \le\
  C\ {e^{-\varepsilon\tau}\over{\la \tau\ra^{r}}},
  \label{eq:2.20}\ee
  for $r<\sigma$. The desired estimate on $S_1^\varepsilon$ 
   then follows by use of (\ref{eq:2.20})  in
  (\ref{eq:integralrep}) and that $\sigma>\sigma_*(n)$.

  It is simple to see that (\ref{eq:2.20}) is
  expected. For, suppose that instead of $B\Pc$  we had $B_0$. Let
  $\omega(k) = \sqrt{m^2+k^2}$, the dispersion relation of $B_0$.
  Then, setting  $L= \left(i\D_{k_j}\omega\right)^{-1}\D_{k_j}$, 
	 and using that 
  $|\nabla_k\omega|\ne0$ on the support of
  $g_\Delta$,
   we get
  \ba
  e^{iB_0t}g_\Delta(B_0) f\ &=&\  \int\ e^{i\omega(k)t}\ e^{ik\cdot x}g_\Delta(k) 
   \hat f(k)\ dk
  \nn\\
  &=&\ t^{-r}\ \int\ L^r\left( e^{i\omega(k)t}\right) \ e^{ik\cdot
  x}\ g_\Delta(k) \hat f(k)\ dk\nn\\
  &=&\ t^{-r}\ \int\ e^{i\omega(k)t} \left(L^{\dagger}\right)^{r} \ 
   \left(e^{ik\cdot x} g_\Delta(k) \hat f(k)\right)\ dk.
   \label{eq:express} \ea
   Estimation of $\la x\ra^{-\sigma} e^{iB_0t}g_\Delta(B_0) f$ in $L^2$ by
   Fourier transform methods, using that 
    $|\nabla \omega|$ is bounded away from zero on the
   support of
	 $g_\Delta$,
  we have that if $\sigma >\max\{ {n\over2},2\}$ and $r<\sigma$, then 
	\be
	\|\la x\ra^{-\sigma}e^{iB_0t}g_\Delta(B_0)f\|_2\ \le\ C\la
	t\ra^{-r}\|\la x\ra^\sigma f\|_2.
	\label{eq:B0variant1}\ee
	Use of this estimate, in the above expression for $S_1^\veps$ (with 
	$B$ replaced by $B_0$) gives, for $l=0,1,2$:
	\be
	\|\la x\ra^{-\sigma}e^{iB_0t}\left(B_0-\Lambda+i0\right)^{-l}\la
	x\ra^{-\sigma}\|_{{\cal B}(L^2)}\le C\la t\ra^{-r+l}.
	\label{eq:B0variant2}\ee

	Two approaches to proving  
	  an estimate of the type (\ref{eq:B0variant1}) and 
	   (\ref{eq:B0variant2}) for $e^{iBt}g_\Delta(B)\Pc$ are as follows:

	 \nit (a) One can "map" the estimate for $B_0$ to that for
	 $B\Pc$ using the wave operator $W_+$. In this approach we would
	 need to derive estimates for $W_+g_\Delta(B_0)$ in weighted $L^2$
	  spaces, or alternatively 

	  \nit (b) One can use the approach based on the "Mourre estimate", 
	   developed in quantum
	  scattering theory. This approach is more in the spirit of energy
	  estimates for partial differential equations and does not require
	  the use of wave operators. 

%%%%%%%%%%%%%%%%%%
Here, we shall follow the latter approach. Time decay estimates like
(\ref{eq:B0variant1}) are a consequence of an approach to {\it minimal velocity
estimates} of Sigal and Soffer \cite{kn:SigSof}, which were proved using
the ideas of Mourre \cite{kn:Mourre}; see also Perry, Sigal \& Simon 
 \cite{kn:PSS}. For our
application to the the operator $e^{-iBt}g_\Delta(B)$, we shall refer to
special cases of  results 
 stated in Skibsted \cite{kn:Skibsted} and Debi\'evre, Hislop \&
Sigal
\cite{kn:DHS}.

We first introduce some definitions and assumptions:.

\nit {\bf (A0)}\ 
 
\ (a) Let  $N\ge2$ and $g_\Delta\in C_0^\infty$, be a smooth
characteristic function which is equal to one on the open interval $\Delta$.

\ (b) Let $H$ and $A$  denote self adjoint operators on a Hilbert
space ${\cal H}$. Assume $H$ is bounded below.
 Let ${\cal D}$ denote the domain of $A$ and  
${\cal D}(H)$ the domain of $H$. Assume ${\cal D}\cap{\cal D}(H)$ is
dense in ${\cal D}(H)$,  and 
 let  $\langle A\rangle\equiv (I+|A|^2)^{1\over2}$.
\medskip

\nit {\bf (A1)}\ Let ${\rm ad}_A^0(H)=H$. For $1\le k\le N_*$,  define
iteratively the commutator form 
\be i^k{\rm ad}_A^k(H)\ =\ i\left[\ i^{k-1}{\rm ad}_A^{k-1}(H)\ ,\ A\ \right]\nn
\ee
 on ${\cal D}\cap{\cal D}(H)$. Assume that for $1\le k\le N_*$, $i^k{\rm
 ad}_A^k(H)$ extends 
to a symmetric operator with domain ${\cal D}(H)$,  
 where $N_*=\left[ N+{3\over2} \right]+1$.
\medskip

\nit {\bf (A2)}\ For $|s|<1$,\  $e^{iAs} : {\cal D}(H)\ \rightarrow
{\cal D}(H)$ and $\sup_{|s|<1}\|\ He^{iAs}\psi\ \|_{\cal H}\ <\ \infty,$
for any $\psi\in {\cal D}(H)$.

\medskip
\nit {\bf (A3)}\ {\it Mourre estimate}: 
\be
g_\Delta (H)\ i[H,A]\ g_\Delta (H)\ \ge \theta\ g_\Delta(H)^2\ 
\label{eq:mourre}
\ee
for some $\theta>0$.
\bigskip

Under the above assumptions, we have, via Theorem 2.4  of
\cite{kn:Skibsted}, the following:

\begin{theo}\  Assume conditions (A0)-(A3). Then,
 for all $\varepsilon_1 >0$ and $t>0$
 \be
  \|\ F\left({A\over t}<\theta\right)\ e^{-iHt}\ g_\Delta (H) 
   \langle A\rangle^{-{N\over2}}\psi\
	\|_{\cal H}
	 \ \le\ C \la t\ra^{ -{N\over2}+\varepsilon_1 }\ \|\ \psi\ \|_{\cal
	 H},
	  \label{eq:mvb}
	   \ee
 where $\theta$ is as in (\ref{eq:mourre}).
 \end{theo}
\medskip

To prove this theorem we set $A(\tau)\ =\ A\ -\ b\tau$, where $\tau\equiv t+1$.
 Then, for any $b<\theta$, we have by
(\ref{eq:mourre}) that the Heisenberg derivative,  
 $DA(\tau)\ \equiv \D_tA(\tau) + i[H,A(\tau)]$, 
satisfies
\be
g_\Delta\ DA(\tau)\ g_\Delta\ =\ 
 g_\Delta\ \left( i[H,A] - b \right) g_\Delta\ \ge \left(\theta\ -\
 b\right)g_\Delta^2.
 \nn \ee
 The result now follows by an application of Theorem 2.4 of  
\cite{kn:Skibsted}.

\nit {\bf Remark:}  The hypotheses of Theorem 2.4 can be relaxed \cite{kn:HSS}
 to the following two conditions:
(i) the operator  $g_\Delta(H)\ {\rm ad}_A^n(H)\ g_\Delta(H)$ can be extended to a
bounded operator on ${\cal H}$ for $n=0,1,2...,\left[{N\over2}+1\right]+1$, and 
 (ii) the Mourre estimate, (\ref{eq:mourre}). 
\bigskip

We shall apply Theorem 2.4 for the case $H = B = \sqrt{-\Delta + m^2
+V(x)}$, $A=\left(x\cdot p\ +\ p\cdot x\right)/2$, $p=-i\nabla_x$ and ${\cal H} = L^2$.  
Before verifying the hypotheses of Theorem 2.4,  
  we show how this
theorem is used  to
derive the  desired estimates on $e^{iBt}g_\Delta(B)$. 
With the application of Theorem 2.4 in mind, we  estimate the 
 norm of $\langle x\rangle^{-\sigma} e^{iBt}g_\Delta(B)$ 
  by decomposing it into the
spectral sets on which $A/t$ is less  than and greater than or equal to
$\theta$. Let $\Delta_1$ denote an interval containing $\Delta$ and denote by
 ${\cal C}$ the operator that associates to a function $f$ its complex
 conjugate ${\overline f}$. Furthermore, note
  that $e^{iBt} \ =\ {\cal C}\ e^{-iBt}\ {\cal
 C}$ and  ${\cal C}\ g(B)\ =\ g(B)\ {\cal C}$, if $g$ is real-valued on
 the spectrum of $B$. Then
 we find,
\ba
 \|\ \la x\ra^{-\sigma}\ e^{iBt}g_\Delta(B)\psi\ \|_2
  \ &=& \|\ \la x\ra^{-\sigma}\ g_{\Delta_1}(B)\ e^{iBt}\ g_\Delta(B)\psi\ \|_2  
  \nn\\
  &=& \|\  \la x\ra^{-\sigma}g_{\Delta_1}(B)  {\cal C}\
   e^{-iBt}\  {\cal C}\ g_\Delta(B)\psi\|_2
	\nn\\
	 \ &=& 
	 \|\  \la x\ra^{-\sigma}g_{\Delta_1}(B)\ \la A\ra^\sigma\cdot
	  \la A\ra^{-\sigma}
	   \ e^{-iBt}\ g_\Delta(B){\overline\psi}\ \|_2\nn\\
		&\le&\
\|\  \la x\ra^{-\sigma}g_{\Delta_1}(B)\ \la A\ra^\sigma\|_{{\cal  B}(L^2)}
\cdot
 \|\ \la A\ra^{-\sigma}
  \ e^{-iBt}\ g_\Delta(B){\overline \psi}\ \|_2\nn\\
  &\le&\ C\ \|\ \la A\ra^{-\sigma}
   \ e^{-iBt}\ g_\Delta(B) {\overline\psi}\ \|_2\nn\\
	 &\le&\ C_1 \|\ F\left({A\over t} <\theta\right)\ \la
	  A\ra^{-\sigma}
	  \ e^{-iBt}\ g_\Delta(B){\overline\psi}\ \|_2\ \nn\\
	   &&\  +\ C_2\|\ F\left({A\over t}\ge \theta\ \right) \la
		A\ra^{-\sigma}
		 \ e^{-iBt}\ g_\Delta(B){\overline\psi}\ \|_2
		  \label{eq:split}
		   \ea
We have used here that, with $A$ defined as above, the $L^2$ operator norm
of $\la x\ra^{-\sigma}g_{\Delta_1}(B) \la A\ra^\sigma$ is bounded; see
\cite{kn:PSS}.
By  Theorem 2.4, the first term on the right hand side of (\ref{eq:split}) can
be 
estimated from above by:
\ba 
&&C_1 \|\ F\left({A\over t} <\theta\right)\ \la A\ra^{-\sigma}
	 \ e^{-iBt}\ g_\Delta(B) g_{\Delta_1}(B) {\overline\psi}\  \|_2\
	 \le\nn\\
&&\ \ \ \  C_1 C\la t\ra^{-{N\over2}+\varepsilon_1}
		\|\ \la A\ra^{N\over2}g_{\Delta_1}(B)\la x\ra^{-{N\over2}}\|_{{\cal
						B}(L^2)}\ 
						 \|\la x\ra^{N\over2}\psi\|_2.\label{eq:term1}
						  \ea
		 Since $\theta t>0$, the second term is bounded as follows:
\be
C_2\|\ F\left({A\over t}\ge \theta\ \right) \la
		A\ra^{-\sigma}
 \ e^{-iBt}\ g_\Delta(B) {\overline \psi}\ \|_2\  
		  \le\ C\la t\ra^{-\sigma}\
		 \|\psi\|_2.
 \label{eq:term2}\ee
	  From (\ref{eq:split}), (\ref{eq:term1}) and (\ref{eq:term2}), we have
	 \be
  \|\la x\ra^{-\sigma}e^{iBt}g_\Delta(B)\psi\|_2\ \le\
   C\left( \la t\ra^{-\sigma}\ +\ \la
   t\ra^{-{N\over2}+\varepsilon_1}\right)\ 
  \|\la x\ra^{N\over2} \psi\|_2.
		   \label{eq:S1nearsingularity}
   \ee
  Use of (\ref{eq:S1nearsingularity}) in a direct
	   estimation of
(\ref{eq:integralrep}) gives:
	 \be
  \|\la x\ra^{-\sigma} S^\veps_1(t)\la
 x\ra^{-\sigma}\|_{{\cal B}(L^2)}\le
  C\ \left(\ \la t\ra^{-{N\over2}+2+\varepsilon_1}\ +\ \la t\ra^{-\sigma +2} \right).
  \label{eq:S1nearsingularityA}\ee
We make the final choice of $N$ and $\sigma$ after estimation of
$S_2^\veps(t)$.
\bigskip

To complete the estimation of $S_1^\veps(t)$, it
 remains to verify hypotheses (A0)-(A3).
These hypotheses are known to hold for the operator 
$H=B^2=-\Delta+V +m^2$; see \cite{kn:CFKS}. Hypothesis (A0) holds $H=B$ with domain
$W^{1,2}(\R^n)$. Clearly hypothesis (A2) holds for $H=B$ since it holds for 
$H=B^2$. Hypothesis
(A1) can be reduced to its verification for $B^2$ 
using the Kato  square root formula \cite{kn:RS1}: for any $\psi\in{\cal D}(B^2)$:
\be
B\psi =\ \pi^{-1}\ \int_0^\infty\ w^{-1/2}\ B^2\ (B^2 + w)^{-1}\ \psi\ dw.
\label{eq:Katosqrt1}
\ee

To verify (A1) for the case $k=1$ we must show that $[A,B]\ B^{-1}$
is a bounded operator on $L^2$.
Using (\ref{eq:Katosqrt1}) we have:
\be
[A,B]\ B^{-1}\ =\ \pi^{-1}\ 
 \int_0^\infty\ w^{1\over2}\ (B^2+w)^{-1}\ [A,B^2]\
 B^{-1}\ (B^2+w)^{-1}\ dw 
\label{eq:AB}
\ee
Since 
\be
\left[ B^2,iA\right]\ =\ 2B^2\ -x\cdot\nabla V\ -2\left( V\ +\ m^2 \right),
\nn\ee
we have
that  
\be
[B^2,iA]\ B^{-1}\ =\ 2 B\ -\ \left( x\cdot\nabla V + 2V
\right)
\ B^{-1}\
-\ 2m^2B^{-1}.\nn\ee
Substitution into (\ref{eq:AB}) we get:
\ba
[A,B]\ B^{-1}\ &=&\ {2\over\pi}\ \int_0^\infty\ w^{1\over2}\ B\ (B^2+w)^{-2}\
dw\nn\\
\ &-&\ {1\over\pi}\ \int_0^\infty\ (B^2+w)^{-1}\ \left(x\cdot\nabla V\ +2V -
2m^2\right)
 B^{-1} (B^2+w)^{-1}\ dw\nn\\
 &\equiv&\ J_1\ +\ J_2.
\ea
The term $J_1$ can be rewritten as
\be J_1\ =\ -{2\over\pi}\ \int_0^\infty\ w^{1\over2}\ {d\over dw}
(B^2+w)^{-1}\ dw\ B\nn\\
=\ I,\nn\ee 
which follows from integration by parts and the formula (\ref{eq:Katosqrt1}) 
 with $\psi$ replaced by $B^{-1}\psi$.
Using that $\|(B^2+w)^{-1}\|\le (\Omega^2+w)^{-1}$,  
\be
\|\ \left(x\cdot\nabla V\ +2V - 2m^2\right) B^{-1}\ \|\ \le\ \|x\cdot\nabla
V\ +2V - 2m^2\|_\infty\ 
\|B^{-1}\|_{{\cal B}(L^2)},\nn\ee
and  the hypotheses on $V$ we have that 
for some constant $C$ dependent on $V$,
\be
|J_2|\ \le\ C\int_0^\infty\ w^{1\over2}\ (\Omega^2+w)^{-2}\ dw\  <\ \infty.
\nn\ee
The higher order commutators are handled in a similar manner; they are even
simpler because each successive commutator results in an extra factor of
$(B^2+w)^{-1}$.

This leaves us with verification of the Mourre estimate (A3) for $H=B$. Proposition
2.2 of \cite{kn:DHS} says that under hypotheses on $B^2$, verified in \cite{kn:CFKS}, 
that (A3) holds for $B$ as well.

\bigskip
\centerline{\it Estimation of  $S_2$:}
\medskip
In $S_2(t)$, the energy is localized away from $\Lambda$ so we seek to
use the $W^{k,s}$ estimates of Proposition 2.1. Note that of the
cases $l=1$ and $l=2$, the case $l=1$ is "worse" because ${\overline g}_\Delta$
localizes the energy away from the singularity at $\Lambda$, and the
$l=1$ term has slower decay for large energy. We therefore carry out the
estimation for $l=1$.  

Let $q^{-1}+p'^{-1}=1/2$ and $p'^{-1} + p^{-1}=1$.
Since $\sigma > {n\over2}$, $\la x\ra^{-\sigma}\in L^q$ and therefore:
\ba
&&\| \la x\ra^{-\sigma}\ S_2^\veps(t)\ \la x\ra^{-\sigma}\ f \|_2
\ \le\ 
 C\|\la x\ra^{-\sigma}\|_{{\cal B}(L^{p'},L^2)}\ 
\|e^{iB_0t}(B_0-\Lambda+i0)^{-1}{\overline g}_\Delta(B_0)\ W_+^* 
 \la x\ra^{-\sigma} f\|_{p'}\nn\\
&& \ \  \le\ C\|e^{iB_0t}B_0^{-1}\|_{{\cal B}(W^{1,p},L^{p'})} \
  \| {\overline g}_\Delta(B_0)(B_0-\Lambda+i0)^{-1} B_0 W_+^* 
   \la x\ra^{-\sigma}\ f \|_{1,p}\nn\\
&&\ \ \le\
 C\|e^{iB_0t}B_0^{-1}\|_{{\cal B}(W^{1,p},L^{p'})} \ 
 \| {\overline g}_\Delta(B_0)(B_0-\Lambda+i0)^{-1}B_0\cdot B_0 W_+^*
	\la x\ra^{-\sigma}\ f \|_p\nn\\
&& \ \ \le\ C\|e^{iB_0t}B_0^{-1}\|_{{\cal B}(W^{1,p},L^{p'})} \ 
 \| {\overline g}_\Delta(B_0)(B_0-\Lambda+i0)^{-1}B_0 \|_{{\cal
                                                     B}(L^p)}
   \|B_0 W_+^* \la x\ra^{-\sigma}\ f \|_p
   \label{eq:S2estimate}
\ea
The three factors in (\ref{eq:S2estimate}) are estimated as follows:

(i) By part (a) of Theorem 2.1,
\be
\|e^{iB_0t}B_0^{-1}\|_{{\cal B}(W^{1,p},L^{p'})}\ \le\ 
 C\ |t|^{-{2n\over n+2}},  
\ee
where $p=2-8/(n+6)$ and $p'= 2 (n+2)/(n-2)$. 

(ii) $\| {\overline g}_\Delta(B_0)(B_0-\Lambda+i0)^{-1}B_0 \|_{{\cal
													 B}(L^p)}$
 is bounded because ${\overline g}_\Delta(\mu)(\mu-\Lambda+i0)^{-1}\mu$
 is a multiplier on $L^p$ \cite{kn:Stein}.

\nit
 Using the boundedness of $W_+$ on $W^{2,p}$ and that
 $\sigma>\max\{ {n\over2},2 \}$, we have 
 \ba
 \|B_0 W_+^* \la x\ra^{-\sigma}\ f \|_p\ &\le&\ \|W_+^* \la
  x\ra^{-\sigma}\ f \|_{1,p} \nn\\
 &\le&\ \| W_+^*\|_{{\cal B}(W^{2,p})}\ \| \la x\ra^{-\sigma}\ f\|_{1,p}\nn\\
 &\le&\ C\| f\|_{1,2}.
 \nn\ea
 Therefore,
 \be
 \| \la x\ra^{-\sigma} S_2^\veps \la x\ra^{-\sigma} \psi\|_2\ \le\ 
 C\la t\ra^{-{2n\over{n+2}}}\ \|\psi\|_{1,2}.
 \label{eq:S2awayfromsingularityA}
 \ee

 We now combine the estimates of $S_j^\veps(t),\ j=1,2$ to complete the proof.
 Combining (\ref{eq:S1nearsingularityA}) and
 (\ref{eq:S2awayfromsingularityA}) and taking $\veps\downarrow 0$ yields
 \be
 \| \la x\ra^{-\sigma}\ e^{iBt}\ \left( B-\Lambda + i0\right)^{-l}\
   \Pc\ \la x\ra^{-\sigma}\ \psi \|_2
 \ \le\
  C\ \left(  \la t\ra^{-{N\over2} + 2 +\varepsilon_1}\ +\ \la t\ra^{-\sigma +2}\ +\
  \la t\ra^{ -{2n\over n+2} }\  
   \right) \|\psi\|_{1,2}.
  \nn\ee
  The estimates (\ref{eq:ld2}) now follows by taking $N$ 
  and $\sigma$ such that 
  \ba
  \sigma\ &\ge&\ 2\ +\ {2n\over n+2}\nn\\
  \sigma\ &>&\ \max\{ {n\over2},2 \}\nn\\
  {N\over2}\ &\ge&\ 2\ +\ {2n\over n+2}\ +\ \veps_1\nn
  \ea
  The constraints on $\sigma$ are implied by the hypothesis $\sigma>\sigma_*(n)$. 
  The remaining constraint holds if $N\ >\ 8$. Since we have applied Theorem 2.4
  in our estimation of $S_1^\veps$, we need that (A1) hold with $N_*=\left[
  N+{3\over2}\right]+1\ge10$.  

  This completes the proof of Proposition 2.2.

\bigskip
\section{Existence theory}
In this section we outline an existence theory for (\ref{eq:nlkg})
with initial conditions $u(x,0) = u_0\in W^{2,2}(\R^3)$ and $\D_tu(x,0) = u_1\in
W^{1,2}(\R^3)$. 
%Since for $f\in W^{1,4}$,
%\be
%\|f\|_8\ \le\ C\ \|\nabla f\|_4^{3\over8}\ \| f\|_4^{5\over8},\ \label{eq:L8}
%\ee
%one it is natural to attempt an existence theory in $W^{1,4}$.
We first reformulate the initial value problem  and then introduce the 
 hypotheses 
 on the operator $H$.
 Regarding (\ref{eq:nlkg})
as a perturbation of a linear constant coefficient equation, we first
write the initial value problem  as:
\ba
\D_t^2 u\ +\ B_0^2 u\ &=&\ -Vu \ +\ \lambda f(u).\label{eq:ivpnlkg1}\\
u(x,0)\ &=&\ u_0(x)\nn\\
\D_tu(x,0)\ &=&\ u_1(x)\nn
\ea
Let
\ba
\bu\ &=&\  \pmatrix{ u\cr v},\ \ U(t)\ =\ \pmatrix{ \cos(B_0t) & \sin(B_0t)/B_0\cr
											  -B_0\sin(B_0t) &\ \cos(B_0t) },\nn\\
\bu_0\ &=&\ \pmatrix{ u_0 \cr u_1},\ \  
{\bf F}(\bu)\ =\ \pmatrix{0 \cr -V u\ +\ \lambda f(u)}.\nn 
\ea
Then, the initial value problem can be reformulated as a system of first order
equations:
\be
\D_t\bu\ =\ \pmatrix{0&1\cr-B_0^2&0}\bu\ +\ {\bf F}(\bu).\nn 
\ee

We now follow the strategy of reformulating  the problem of finding a 
 solution of the initial value problem 
 as the problem of finding a fixed
point $\bu$ of an appropriate mapping. In particular, we seek $\bu$ in the space
\be
\bXz\ =\ W^{2,2}(\R^3)\ \times\ W^{1,2}(\R^3)
\label{eq:Xzspace}
\ee
 satisfying  
\be
{\cal A}\ \bu \ =\ \bu, \label{eq:integral-equation}
\ee
where 
\be
{\cal A}\bu(t)\ \equiv\ U(t)\bu_0\ +\ \int_0^t U(t-s)\ F(\bu)\ ds. \label{eq:Adef}
\ee
\medskip

 \noindent 
 For a discussion of
 the existence and low energy scattering for the case $V\equiv 0$, see
 \cite{kn:GV}, \cite{kn:strauss}, \cite{kn:Kl} and references
 cited therein.

\bigskip
\nit\ \ {\bf Hypotheses of H\ =\ $-\Delta +V$} 

\nit {\bf (H1)}\ $V\in  W^{1,\infty}$

\nit {\bf (H2)}\ $ H=B^2 $ is a positive and self-adjoint operator

\nit {\bf (H3)}\ The  semi-infinite interval, $[m^2,\infty)$, consists of
absolutely continuous spectrum of $H$.

\nit {\bf (H4)}\  $H$ has exactly
  one (simple)
   eigenvalue $\Omega^2$ satisfying $0\ <\  \Omega^2\ <\  m^2$,   
   with corresponding eigenfunction $\varphi\in
  L^2,\
  ||\varphi||_2=1.$

\nit {\bf (N)}\  $f(u)\ =\ u^3\ +\ f_4(u),\  f_4(u)\ =\ {\cal O}(u^4)$
 and $f(u)$ is smooth in a neighborhood of $u=0$.

These hypotheses are by no means the least stringent, but are
sufficient for the present purposes.

\begin{theo} ({\bf Local existence theory}) 

\nit
Consider the Cauchy problem for the nonlinear Klein Gordon equation 
 (\ref{eq:nlkg}) with
initial data, $\bu_0$ of class $\bXz$. 

\item{(a)} There exists strictly positive
numbers,
 $T_{max}$ and $T^{max}$ (forward and backward 
maximal times of existence) which depend on the  $\bXz$ norm of the
initial data,  such that the initial value problem has a
unique solution of class 
 $C^0\left((-T^{max},+T_{max}); \bXz\right)$ in the sense of the integral
 equation (\ref{eq:integral-equation}).

\item{(b)} For $t\in (-T^{max},+T_{max})$ conservation of
energy holds: ${\cal E}[u(\cdot,t), \D_t u(\cdot,t)]\ =\ 
         {\cal E}[u_0, u_1]$, where ${\cal E}$ is defined in equation
		  (\ref{eq:energy}).

\item{(c)} Either $T_{max}$ is finite or $T_{max}$ is infinite. If
$T_{max}$ is finite, then
\be
\lim_{t \uparrow T_{max}}\ \| \bu(\cdot,t)\|_{\bXz}\ =\ \infty.
\label{eq:Xz-controls}
\ee
The analogous statement holds with $T_{max}$ replaced by $T^{max}$ and 
$\lim_{t\uparrow T_{max}}$ replaced by $\lim_{t\downarrow -T^{max}}$ .
\end{theo}

We now sketch a proof of Theorem 3.1. We restrict our attention to the case
$t\ge0$; the proof is identical for $t\le0$. Furthermore,
  we shall, for simplicity, consider the
case of the cubic nonlinearity: $f(u)=u^3$. 
 Using hypothesis (H1) on $V$, 
  it is simple to show, 
   using the estimates of Corollary 2.1,  
   that for $T>0$ sufficiently
 small and depending essentially on the $\bXz$ norm of $\bu_0$,  
  the mapping 
  ${\cal A}$ maps  a closed ball in $C^0\left( [0,T);\bXz\right)$
 to itself, and is a strict contraction. This ensures the existence of
 a unique fixed point, which is a $C^0\left( [0,T);\bXz\right)$ solution 
  of the initial value problem (\ref{eq:ivpnlkg1})
 in the sense of the integral equation (\ref{eq:integral-equation}). 
That the $\bXz$ norm must blow up if $T_{max}$ is finite is a standard
continuation argument based on the fixed point proof of part (a).

In section 7, we study the asymptotic behavior of solutions as
$t\to\pm\infty$. A consequence of this analysis and the argument below 
 is that if $\bu_0$ satisfies
the more stringent hypothesis that  the norm $\|\bu_0\|_{\bX}$ is sufficiently
small, where 
\be
 {\bX}\  \equiv\ \left( W^{2,2}\cap W^{2,1}\right)\ \times 
 \ \left(W^{1,2}\cap
W^{1,1} \right),\label{eq:bX}
\ee
 then $T_{max}=T^{max}=\infty$ and the solution decays to zero
as dilineated in the statement of Theorem 1.1.

We reason as follows. A unique solution, $\bu(t)$,  
 exists locally
in time and is continuous in $t$ with values in $\bXz$. 
 It follows that for $|t|<T$, the $W^{1,4}$ and $L^8$ norms
of $u(\cdot,t)$ are finite. Moreover, the solution satisfies energy
conservation and therefore $\|u(t)\|_{W^{1,2}}+\|v(t)\|_2$ is 
 bounded uniformly  by a
constant which is independent of $T$, and depends only on $\|u_0\|_{W^{1,2}}+
 \|v_0\|_2$.
To ensure that the quantities estimated continue to be well defined and satisfy
the {\it 'a priori} estimates of section 7 it suffices to show that 
$\|\bu(t)\|_{\bXz}$ remains bounded as $t\uparrow T$.
A direct and  elementary estimation of the integral equation
(\ref{eq:integral-equation})  yields the estimate: 
\ba
\|\bu(t)\|_{\bXz}\ &\le& C\left(\|u_0\|_{W^{2,2}}\ +\ \|u_1\|_{W^{1,2}}\right)\ \nn\\
&+&\
C\left(\|V\|_{W^{1,\infty}},\lambda\right)\ 
 \int_0^t\ \left( \|u(s)\|_{W^{1,2}}\ +\ \|u(s)\|_{W^{1,2}}^3\ +\ 
 \|u^2(s)\nabla u(s)\|_2\ \right) ds.\nn\\
 &&\label{eq:H2ap}
 \ea
The main tool used in obtaining (\ref{eq:H2ap}) is the bound
$\|E_1^{(0)}(t)\D_i\|_{{\cal B}(L^2)}\le m^{-1}$.  

A simple consequence of the {\it 'a priori} bound (\ref{eq:aprioriest}) and
the decomposition of $u(t,\cdot)$ is
that 
\be
\|\nabla u(t)\|_4\  +\ \| u(t)\|_8\ \le\  {\tilde C},
\ |t|<T.
\label{eq:bound}
\ee
 The constant $\tilde C$, depends on the norm $\| \bu_0\|_{\bX}$.
 The first two terms in the integrand of (\ref{eq:H2ap}) are uniformly bounded
 by conservation of energy. 
Furthermore, this energy bound together with (\ref{eq:bound}) 
 implies a bound on the last term in the integrand
of the estimate (\ref{eq:H2ap}). It follows that $\|\bu(t)\|_{\bXz}$ remains
bounded as $t\uparrow T$. Therefore, given the estimates of section 7 we 
 have $T_{max} =\infty$, and the decay of solutions.

A further consequence of the proof is:
\begin{cor}
Under the hypotheses of Theorem 1.1, the solution of the initial value problem
exists globally in $W^{2,2}(\R^3)$ and satisfies the estimate:
\be \|u(t)\|_{W^{2,2}}\le C\ \la t\ra.\nn\ee
\end{cor}

\bigskip
\section{ Isolation of the key resonances and formulation as coupled
finite and infinite dimensional dynamical system}
%
%%%%%%%%%%%%%%%%%%%%%%%%%%%%%%%%%%%%%%%

Using the notation (\ref{eq:Bdef}), the initial value problem for 
 (\ref{eq:nlkg}) can be rewritten as
\ba
&&\partial^2_t\ u\ +\ B^2\ u\ =\ \lambda\ f(u),\ \ f(u)=u^3\label{eq:nlkg1} \\
&&u(x,0)\ =\ u_0(x),\  \partial_t\ u(x,0)\ =\ u_1(x).\nn
\ea
For small amplitude solutions,
 it is natural to decompose the solution as follows:
 \ba
 u(x,t)\ &=&\ a(t)\varphi (x)+\eta (x,t),
 \label{eq:ansatz}\\
 \left( \eta(\cdot,t),\varphi\right)\ &=&\ 0 \quad \hbox{\rm for
 all}\ \ t\
 .
 \label{eq:orthog}
 \ea
Substitution of (\ref{eq:ansatz}) into (\ref{eq:nlkg1}) gives
\be
a''\varphi\  +\ \partial^2_t\eta\ +\ \Omega^2 a\varphi\ +\  B^2\eta\
 =\ \lambda\  f(a\varphi\ +\ \eta)
 \label{eq:a1}
 \ee
 We now implement (\ref{eq:orthog}). Taking the inner product of
  (\ref{eq:a1}) with $\varphi$ gives 
  \be
   a''\ +\ \Omega^2 a\ =\ \lambda(\varphi,f(a\varphi+\eta)).
   \label{eq:aeqn}
   \ee
   Let $\Pc$ denote the projection onto the continuous spectral part
   of
   $B^2$, i.e.
   \be
   \Pc\ v\equiv
   v-(\varphi,v)\varphi. \label{eq:Pc}
   \ee
	Then, since $\eta = \Pc\eta$, we have 
	\be
	\partial^2_t\eta\ +\ B^2\eta=\lambda \Pc f(a\varphi+\eta)
	\label{eq:etaeqn}
	\ee

Equations (\ref{eq:aeqn})-(\ref{eq:etaeqn}) comprise a coupled dynamical
system
for the bound state and continuous
spectral components (relative to $H=B^2$) of the solution $u$.
The initial conditions for this system are given by
(\ref{eq:aetadata}).

Expansion of the cubic terms in (\ref{eq:aeqn}-\ref{eq:etaeqn})
  gives the system
\ba
 a''\ +\ \Omega^2a\ &=&\ \lambda\ 
 \left[\ a^3\int\varphi^4\ +\ 3a^2\int\varphi^3\eta\ +\ 3a\int
\varphi^2\eta^2\ +\ \int\varphi\eta^3\ \right]\
\label{eq:aeqn2}\\
\D_t^2\ \eta\ +\ B^2\ \eta\ &=&\ \lambda \Pc\ \left( 
 a^3\varphi^3\ +\ 3a^2\varphi^2\eta
\ +\ 3a\varphi\eta^2\ +\ \eta^3\ \right) ,\label{eq:etaeqn2}
\ea
with initial conditions
\ba
a(0)&=&(\varphi,u_0)\ ,\quad a'(0)=(\varphi,u_1)\ \nn\\ 
\eta(x,0)\ &=&\  \Pc u_0\ ,\quad\partial_t\eta(x,0)\ =\ \Pc u_1
\label{eq:data}
\ea

We now locate the source of the resonance. Equation (\ref{eq:aeqn2}) 
 has a homogeneous solution which  
oscillates with frequencies $\pm\Omega$. Thus, to leading order, $\eta$ solves a driven wave
equation containing the driving frequencies $\pm 3\Omega$. If $9\Omega^2>m^2$, then by $(H3)$, we expect a
resonant interaction with radiation modes of energy 
 $3\Omega\in\sigma_{cont}(H)$.

This resonant part of $\eta$, has its dominant effect on the $a$-oscillator 
 in the term of (\ref{eq:aeqn2}),
which is linear in $\eta$. Our goal is to derive from 
 (\ref{eq:aeqn2}-\ref{eq:etaeqn2}) an equivalent dynamical system
which is of the type described in the introduction, but which is corrected by terms which decay sufficiently 
rapidly with time and which can therefore be  treated perturbatively.

We first write
\be
\eta=\eta_1+\eta_2+\eta_3\ ,
\label{eq:eta123}
\ee
where $\eta_1(t)$ satisfies the {\it linear dynamics} with the given initial data:
\be
\partial^2_t\ \eta_1\ +\ B^2\eta_1\ =\ 0\ ,\quad\eta_1(x,0)\ =\ \Pc u_0\ ,\ 
  \partial_t\ \eta_1(x,0)\ =\ \Pc\ u_1,\
\label{eq:eta1eqn}
\ee
and $\eta_2(t)$ is the leading order {\it response}
\be
\partial^2_t\ \eta_2\ +\ B^2\eta_2\ =\ \lambda\  a^3\ \Pc\ \varphi^3;\ 
 \eta_2(x,0)=0,\
\partial_t\ \eta_2(x,0)=0\ .
\label{eq:eta2eqn}
\ee
Equation (\ref{eq:aeqn2}) can now be written in an expanded form.
\ba
 a''\  +\ \Omega^2\ a\ &=&\  \lambda\left[\ 
a^3\int\varphi^4+3a^2\int\varphi^3\ \left(\ \eta_1+\eta_2+\eta_3\ \right)
 +3a\int\varphi^2
\eta^2+\int\varphi
\eta^3\ \right]\nn\\
&\equiv&\ F(a,\eta)
%\eqno(4.12)$$
\label{eq:aeqn3}
\ea
We expect the function $a(t)$ to consist of "fast oscillations",
coming from the natural frequency $\Omega$ and its nonlinearly generated
harmonics, and slow variations due to its small amplitude. 
We  next extract from  $a(t)$ the dominant "fast" oscillations of frequency
$\Omega$:
\be
a(t)\ =\ A\ e^{i\Omega t}+\overline A\  e^{-i\Omega t}\ .
%\eqno(4.13)$$
\label{eq:aA1}
\ee
We then substitute (\ref{eq:aA1}) into (\ref{eq:aeqn3}) and impose the
constraint:
\be
A' e^{i\Omega t}\ +\ {\overline A}'e^{-i\Omega t}\ =\ 0
\nn\ee

Equation (\ref{eq:aeqn3}) then is reduced to the first order equation: 
\be
 A'\ =\ (2i\Omega)^{-1}\  e^{-i\Omega t}\  F(a,\eta)\ ,
%\(4.14)$$
\label{eq:Aeqn}
\ee
where $F(a,\eta)=F(A,\overline A,\eta,t)$. 
 From (\ref{eq:aeqn3}) we have that $F(a,\eta)$ is the sum of the following terms: 
\ba
F_1(a)& =&\ \lambda a^3\int\varphi^4\nn\\
F_2(a,\eta_j)&=&\ 3\lambda a^2\int\varphi^3\eta_j\ , \quad j=1,2,3\nn\\
F_3(a,\eta)&=&\ 3\lambda a\int\varphi^2\eta^2\nn\\
F_4(\eta)&=&\ \lambda\int\varphi\eta^3\ .\ \label{eq:F1234}
\ea

The remainder of this section is primarily devoted to an (involved)
  expansion of
 $F_2(a,\eta_2)$. The terms of this expansion are of several types:
 (a) the resonant damping term, (b) terms of the same order
whose net  
 effect is a nonlinear phase correction and (c) higher order
terms which are to be treated perturbatively in the asymptotic analysis
as $t\to\pm\infty$.
\smallskip

\centerline{\it Computation of $F_2(a,\eta_2)$}

We first break $\eta_2$ into a part containing the key resonance $\eta^r_2$ and nonresonant part
$\eta^{nr}_2$.

\smallskip
\begin{prop}
$$\eta_2=\eta^r_2 +\eta^{nr}_2\ ,\label{eq:4.16}$$
where
\be
\eta^r_2={\lambda\over 2iB} e^{iBt}\int^t_0 e^{-is(B-3\Omega)} A^3(s)\
ds\  \Pc\varphi^3\
,\label{eq:eta-r2}
\ee
and
\ba
\eta^{nr}_2&=&\ {\lambda\over 2iB} \e{iBt}\bigg\lbrack 3\i \e{-is(B-\Omega)}
A^2(s)\overline A(s)ds\nn\\
&+&\ 3\i \e{-is(B+\Omega)}\ \overline A^2(s)A(s)\ ds\nn\\
&+&\ \i\e{-is(B+3\Omega)}\ \overline A^3(s)\ ds\bigg\rbrack\  \Pc\varphi^3\nn\\
&-&\ {\lambda\over 2iB} \e{-iBt}\bigg\lbrack\i \e{is(B+3\Omega)}\ 
A^3(s)\ ds\nn\\ 
&+&\ 3\i \e{is(B+\Omega)}\ A^2(s)\overline A(s)\ ds+3\i  \e{is(B-\Omega)}\overline A^2(s) A(s)\ ds\nn\\ 
&+&\ \i\e{is(B-3\Omega)}\ \overline A^3(s)\ ds\bigg\rbrack\  \Pc\varphi^3\nn\\
&\equiv&\ \sum^7_{j=1} \eta^{nr}_{2j}\nn\\ \
%\label{eq:4.18}
\label{eq:eta-nr2}
\ea
\end{prop}
The superscripts $r$ and $nr$ denote respectively a resonant contribution 
 and nonresonant
contribution.
\medskip

\noindent{\it Proof:} 
The solution of (\ref{eq:eta2eqn}) can be expressed, using the
variation of constants formula, as:
\be
\eta_2\ =\ \lambda\i{\sin B(t-s)\over B} a^3(s)\ ds\ \Pc\varphi^3,\nn
 \ee
Substitution of (\ref{eq:aA1}) for $a(s)$ and using the expansion
of $\sin\left(B(t-s)\right)$ in terms of the operators $\exp\left(\pm iB(t-s)\right)$ 
leads to an expression for $\eta_2$ which is a sum of eight terms. The
term 
$\eta_2^r$, as defined above, is anticipated to be the most important.
The other seven terms are lumped together in $\eta_2^{nr}$.
\medskip

We now focus on $\eta_2^r$. 
In order to study $\eta^r_2$ near the resonant point $3\Omega$ in the continuous spectrum of
$B$, we first introduce  a regularization of $\eta^r_2$. For 
 $\veps>0$, let 
\be
\eta^r_{2\veps}\equiv{\lab}\e{iBt}\i\e{-is(B-3\Omega+i\veps)} A^3(s)\
ds\ \Pc\varphi^3\
% \label{eq:4.19}$$
\label{eq:eta-2eps}
\ee
Note that $\eta^r_2=\lim_{\veps\to0}\eta^r_{2\veps}$.
The following result, proved using integration by parts, isolates the
key (local in $t$) resonant term.
\smallskip

\begin{prop} For $\veps\ge0$,  
\ba
\eta^r_{2\veps}&=&\ {\lambda\over 2} \Big\lbrack B(B-3\Omega+i\veps)
 \Big\rbrack^{-1}
\e{3i\Omega t}A^3(t)\ e^{\veps t} \Pc\varphi^3\nn\\
&-&\ {\lambda\over 2}A_0^3 \Big\lbrack B(B-3\Omega+i\veps)\Big\rbrack^{-1}
\e{iB t}\ \Pc\varphi^3\nn\\
&-&\ {3\over 2}\lambda  \Big\lbrack B(B-3\Omega+i\veps)\Big\rbrack^{-1}
\e{iB t}\i \e{-is(B-3\Omega +i\veps )}A^2A'\ ds\ \Pc\varphi^3\nn\\
&=&\ \eta^r_{*\veps}+\eta^{nr1}_{*\veps}+\eta^{nr2}_{*\veps}
 \label{eq:eta-2eps1}
\ea
\end{prop}

\nit{\bf Remark:} The choice $+i\veps$ in (\ref{eq:eta-2eps}) 
 is motivated by the fact that the operator \
\be (B-3\Omega+i0)^{-1}e^{iBt} \nn\ee
satisfies appropriate
 decay estimates as $t\rightarrow \infty$; see Proposition 2.1. 
  It follows that
the limit as $\veps\to0^+$ of the last two terms in (\ref{eq:eta-2eps1})
  decay in time like $\jt^{-1-\alpha}(\alpha>0)$; see section 7.

We now use the above computation to obtain an expression for $F_2(a,\eta_2)$. First, from
Proposition 4.1 we have
\ba
F_2(a,\eta_2)\ &=&\ F_2(a,\eta^r_2)+F_2(a,\eta^{nr}_2)\nn\\
   &=&\ \lim_{\veps\to0} F_2(a,\eta^r_{*\veps})\ +\ 
          \lim_{\veps\to0} F_2(a,\eta^{nr1}_{*\veps} +
           \eta^{nr2}_{*\veps})+F_2(a,\eta_2^{nr})\ ,\nn\\
   &\equiv&\  F_2(a,\eta^r_*)\ +\ F_2(a,\eta^{nr1}_*\ +\ \eta^{nr2}_*)\
+\ F_2(a,\eta_2^{nr})
 \label{eq:F2}
\ea
where $F_2(a,\cdot)$ is defined in (\ref{eq:F1234}). We begin with 
 the contribution to (\ref{eq:Aeqn}) coming from
$F_2(a,\eta^r_*)$. What follows now is a detailed expansion of the term
  $F_2(a,\eta^r_*)$ and
$F_2(a,\eta_2^{nr})$. The terms $F_2(a,\eta_*^{nr1}+\eta_*^{nr2})$ can be 
 treated
perturbatively by estimation of its magnitude; see section 5.
\smallskip

\centerline{\it Computation of $F_2(a,\eta^r_*)$}

Let 
\ba
\Lambda\ &\equiv&\
  \lim_{\veps\rightarrow 0}\left(\ \Pc\varphi^3, {1\over B}\ {B-3\Omega\over
(B-3\Omega)^2+\veps^2} \Pc\varphi^3\ \right)
\label{eq:Lambda}\\
         &=&\ \left(\ \Pc\varphi^3, {1\over B} {\rm P.V.} {1\over
		 B-3\Omega}\ \Pc\varphi^3\ \right),\ {\rm and}\ 
\nn\ea
\ba
\Gamma&\equiv&\ \lim_{\veps\rightarrow 0}\ \left(\ {\bf P}_c\varphi^3, 
 {1\over B}\ {\veps\over
(B-3\Omega)^2+\veps^2} \Pc\ \varphi^3\ \right)\nn\\
&=&\ {\pi\over 3\Omega} \left( \Pc\ \varphi^3,\d(B-3\Omega) \Pc\ \varphi^3\right)\nn\\
&=&\ {\pi\over3\Omega}\ \left|\ {\cal F}_c[\varphi^3](3\Omega)\ \right|^2 .  
 \label{eq:Gamma}
\ea
By hypothesis (\ref{eq:nlfgr}), $\Gamma>0$.

We now substitute the expression for $\eta_{*\veps}^r$, given in 
 (\ref{eq:eta-2eps1}) into the definition
of $F_2(a, \eta_{*\veps}^r)$ in (\ref{eq:F1234}). Passage to the limit, 
 $\veps\to0$,  and use of the distributional identity:
 \be
 (x\pm i0)^{-1}\ \equiv\ \lim_{\varepsilon\to0}\ (x\pm
 i\varepsilon)^{-1}\ =\ {\rm P.V.}\ x^{-1}\ \mp i\pi\delta(x)
 \nn\ee
 yields: 
\smallskip

\begin{prop} 
\ba
F_2(a,\eta^r_*)&=&\ \lim_{\veps\rightarrow 0}F_2(a,\eta^r_{*\veps})\nn\\
&=& {3\over 2}\lambda^2 \left(
      \Lambda-i\Gamma\right) \Big\lbrack|A|^4 Ae^{i\Omega t}  
+A^5 e^{5i\Omega t}+2|A|^2 A^3 e^{3i\Omega t}\Big\rbrack\label{eq:4.23}
\ea
\end{prop}

We have completed the evaluation of the first term in (\ref{eq:F2}). \
 To calculate the contribution of
(\ref{eq:4.23}) to the amplitude equation (\ref{eq:Aeqn}) we need only multiply (\ref{eq:4.23}) by 
$(2i\Omega)^{-1}
e^{-i\Omega t}$. This gives
\smallskip
\begin{prop}
\be
(2i\Omega)^{-1}\ e^{-i\Omega t}\ F_2(a,\eta^r_*)\ 
=\ -{3\over 4}{\lambda^2\over \Omega}\ (i\Lambda+\Gamma)\Big\lbrack |A|^4 A+A^5 e^{4i\Omega t}+2|A|^2A^3
e^{2i\Omega t}\Big\rbrack\label{eq:4.24}
\ee
\end{prop}
The term $-{3\over 4}{\lambda^2\over \Omega}\ \Gamma |A|^4 A$ plays the role of a nonlinear damping; it drives
the decay of $A$ and, in turn, that of $\eta$; see the discussion in the introduction.

\medskip

\centerline{\it Computation of $F_2(a,\eta_2^{nr})$:}

We now focus on $F_2(a,\eta_2^{nr})$, the third term in (\ref{eq:F2}).
This requires a rather extensive, expansion of
$\eta_2^{nr}=\sum_j\eta^{nr}_{2j}$; see (\ref{eq:eta-nr2}). 
 From Proposition 4.3, we expect the dominant
terms to be 
$ {\cal O}(|A|^5)$.
 Our approach 
 is now to make explicit all terms which are formally
$ {\cal O}(|A|^5)$ (anticipating that $A'\ =\  {\cal O}(|A|^3)$, $A''\
=\ {\cal O}(|A|^5)$ and $(|A|^2)'\
=\  {\cal O}(|A|^6)$ ) and to treat the remainder
as a perturbation which we shall later estimate to be of higher order. 
 This expansion of
$\eta_2^{nr}$ is presented in the following proposition which we prove 
 using repeated integration by parts. As written, these expressions are
 formal. By the definition of $F_2(a,\eta_2^{nr})$ we require that they
 hold when integrating against a rapidly decaying function, {\it i.e.}
 $\varphi^3$.

\smallskip
\begin{prop} The following expansions hold in ${\cal S}'$:
\ba
\et_{21}&=&\ {\lambda\over 2B(B-\Omega)} |A|^2 A e^{it\Omega}\ \Pc\varphi^3
+{3\lambda\over 2iB (B-\Omega)^2} e^{it\Omega}(|A|^2A)'\ \Pc\varphi^3\nn\\
&+&\ {3\lambda e^{iBt}\over 2B(B-\Omega)}\Big\lbrack
|A_0|^2A_0+(|A|^2A)'\big|_{t=0}\Big\rbrack\ \Pc\varphi^3\nn\\
&-&\ {3\lambda\over 2iB(B-\Omega)^2} e^{iBt}\i e^{-is(B-\Omega)}\ (|A|^2 A)''\
 ds
 \ \Pc\varphi^3\label{eq:4.25}
\ea
\ba
\et_{22}&=&\ {3\lambda\over 2B(\B)}|A|^2\ov e^{-it\Omega} \va +{3\lambda\over
2iB(\B)^2}e^{-it\Omega} (\ov^2 A'+2{\ov}' |A|^2)\va\nn\\
&+&\ e^{iBt}\Big\lbrack-{3\over 2}\ {\lambda\over 2B(\B)}|A_0|^2\ov -{3\lambda\over 2B(\B)}
(|A|^2A)'\big|_{t=0}\Big\rbrack \va\nn\\
&-&\ {3\lambda\over 2iB(\B)^2} e^{iBt}\i e^{-is(\B)}(|A|^3\ov)''(s)\ ds\
 \va
\label{eq:4.26}
\ea
\smallskip
\ba
\et_{23}&=&\ {\lambda\ov (t)^3\over 2B(B+3\Omega)} e^{-3it\Omega}\ \va
+{3\lambda\over 2iB(B+3\Omega)^2} e^{-it3\Omega} (\ov(t)^3)'\ \va\nn\\
&-&\ {3\over 2}\lambda{e^{iBt}\over B(B+3\Omega)}\Big\lbrack\ov^3_0+(\ov^3)'\big|_{t=0}\Big\rbrack
\va\nn\\
&-&\ {3\over 2}\ {\lambda\over iB(B+3\Omega)^2}e^{iBt}\i e^{-is (B+3\Omega)}(\ov(s)^2)''\ ds\ \va\nn\\
\label{eq:4.27}
\ea
\smallskip
\ba
\et_{24}&=&\ {\lambda\over 2B(B+3\Omega)} A^3(t) e^{3it\Omega}\va
-{3\lambda\over 2iB(B+3\Omega)^2} e^{3it\Omega} (A^3)'\ \va\nn\\
&-&\ {3\over 2}\lambda{e^{-iBt}\over B(B+3\Omega)}\Big\lbrack A^3_0+(A^3)'\big|_{t=0}\Big\rbrack
\va\nn\\ 
&+&\ {3\lambda\over 2 iB(B+3\Omega)^2}e^{-iBt}\i e^{is (B+3\Omega)}(A^3(s))''\ 
 ds\  \va\nn\\
\label{eq:4.28}
\ea
\smallskip
\ba
\et_{25}&=&\ {3\lambda\over  2B(B+\Omega)} |A|^2 A\ e^{it\Omega}\ \va
-{3\lambda\over 2iB(B+\Omega)^2} e^{it\Omega} (|A|^2 A)'\va\nn\\
&-&\ {3\over 2}\lambda{e^{-iBt}\over B(B+\Omega)}\Big\lbrack |A_0|^2 A_0+(|A|^2 A)'
\big|_{t=0}\Big\rbrack
\va\nn\\
&+&\ {3\lambda\over 2 iB(B+\Omega)^2}e^{-iBt}\i e^{is (B+\Omega)}(|A|^2 A)''\
ds\ \va\nn\\
\label{eq:4.29}
\ea
\smallskip
\ba
\et_{26}&=&\ {3\lambda\over  2B(B-\Omega)} |A|^2 \ov \ e^{-it\Omega}\ \va
-{3\lambda\over 2iB(B-\Omega)^2} e^{-it\Omega} (|A|^2 \ov)'\ \va\nn\\
&-&\ {3\over 2}\lambda{e^{-iBt}\over B(B-\Omega)}\Big\lbrack |A_0|^2 \ov_0+(|A|^2 \ov)'
\big|_{t=0}\Big\rbrack\ \va\nn\\ 
&+&\ {3\over 2 i}\lambda {1\over B(B-\Omega)^2}e^{iBt}\i e^{is(B-\Omega)}(|A|^2  \ov )''\ ds\  \va\nn\\ 
\label{eq:4.30}
\ea
\smallskip
\ba
\et_{27}&=&\ {\lambda\ov^3(t)\ e^{-3it\Omega}\over  2B(B-3\Omega-i0)}\ \va
-{3\lambda\over 2iB(B-3\Omega -i0)^2} e^{-3it\Omega}(\ov^3)'\ \va\nn\\
&-&\ {3\over 2}\lambda {e^{-iBt}\over B (B-3\Omega - i0)} 
\Big\lbrack\ov_0^3+(\ov^3)'\big|_{t=0}\Big\rbrack\ \va \nn\\
&+& {3\lambda\over
2iB(B-3\Omega-i0)^2} e^{-iBt}\i e^{is(B-3\Omega)} (\ov^3)''ds\ \va
\label{eq:4.31}
\ea
\end{prop}
Recall that our goal is to elucidate the structure of the amplitude
equation: $A'\ =\ (2i\Omega)^{-1}e^{-i\Omega t}F$, in (\ref{eq:Aeqn}),
where we first focused on the contribution: $(2i\Omega)^{-1}e^{-i\Omega
t}F_2$. From (\ref{eq:F2}) and Proposition 4.4 we see now that we need
to obtain convenient expressions for 
\ba
(2i\Omega)^{-1} e^{-i\Omega t} F_2(a,\et_2)\ &=&\  
 \sum^7_{j=1}(2i\Omega)^{-1} e^{-i\Omega t}\ F_2(a,\et_{2j})\nn\\
&=&\ 3\lambda(2i\Omega)^{-1} a^2 e^{-i\Omega t}\sum^7_{j=1}\int\varphi^3\et_{2j}\
.\label{eq:4.32}
\ea
Each of the seven terms contributing to $\eta_2^{nr}$ 
 is expressed as a part which is ${\cal O}(|A|^5)$ 
 plus an error term
which is estimated in magnitude in section 5.
The $ {\cal O}(|A|^5)$ part and error terms are displayed 
 in the following two propositions.
\smallskip

\begin{prop}
 Let 
\be
\rho(\zeta)\equiv(\Pc\varphi^3, B^{-1}(B-\zeta)^{-1}\Pc\varphi^3).
\label{eq:rhoeqn}
\ee
Then, 
\ba
&&(2i\Omega)^{-1} e^{-i\Omega t} F_2(a,\et_{21})\nn\\
&=&\ {\lambda^2\over\Omega} \rho(\Omega) \Big\lbrack {3\over 4i} |A|^2A^3 e^{2i\Omega t} +{3\over
2i} |A|^4A+ {3\over 4i} |A|^4\ov e^{-2i\Omega t}\Big\rbrack
+E^{nr}_{21}\label{eq:4.33}
\ea
\smallskip
\ba
&&(2i\Omega)^{-1} e^{-i\Omega t} F_2(a,\et_{22})\nn\\
&=&\ {\lambda^2\over\Omega}\rho(-\Omega) \Big\lbrack{9\over 4i}|A|^4A+{9\over 2i}|A|^4\ov
e^{-2i\Omega t}+{9\over 4i}|A|^2\ov^3 e^{-4i\Omega t}\Big\rbrack
+E^{nr}_{22}\label{eq:4.34}
\ea
\smallskip
\ba
&&(2i\Omega)^{-1} e^{-i\Omega t} F_2(a,\et_{23})\nn\\
&=&\ {\lambda^2\over\Omega}\rho(-3\Omega) \Big\lbrack{3\over 4i}|A|^4\ov e^{-2i\Omega t}+{3\over 2}
|A|^2\ov^3
e^{-4i\Omega t}+{3\over 4}\ov^5 e^{-6i\Omega t}\Big\rbrack
+E^{nr}_{23}\label{eq:4.35}
\ea
\smallskip
\ba
&&(2i\Omega)^{-1} e^{-i\Omega t} F_2(a,\et_{24})\nn\\
&=&\ {\lambda^2\over\Omega}\rho(-3\Omega) \Big\lbrack{3\over 4}A^5 e^{4i\Omega t}+{3\over 4}
A^3 |A|^2
e^{2i\Omega t}+{3\over 4}|A|^4 A \Big\rbrack
+E^{nr}_{24}\label{eq:4.36}
\ea
\smallskip
\ba
&&(2i\Omega)^{-1} e^{-i\Omega t} F_2(a,\et_{25})\nn\\
&=&\ {\lambda^2\over\Omega}\rho(-\Omega) \Big\lbrack{9\over 4i}|A|^2 A^3 e^{2i\Omega t}+{9\over 2i}
|A|^4 A+{9\over 4i}|A|^4\ov
e^{-2i\Omega t} \Big\rbrack
+E^{nr}_{25}\label{eq:4.37}
\ea
\smallskip
\ba
&&(2i\Omega)^{-1} e^{-i\Omega t} F_2(a,\et_{26})\nn\\
&=&\ {\lambda^2\over\Omega}\rho(\Omega) \Big\lbrack{9\over 4i}|A|^4 A +{9\over 2i}|A|^4\ov
e^{-2i\Omega t}+{9\over 4i}
|A|^2 \ov^3
e^{-4i\Omega t} \Big\rbrack
+E^{nr}_{26}\label{eq:4.38}
\ea

\smallskip
\ba
&&(2i\Omega)^{-1} e^{-i\Omega t} F_2(a,\et_{27})\nn\\
&=&\ {\lambda^2\over\Omega}\rho(3\Omega +i0) \Big\lbrack{3\over 4i}|A|^4 \ov 
e^{-2i\Omega t}+{3\over 2i}
|A|^2 \ov^3
e^{-4i\Omega t} +\ov^5 e^{-6i\Omega t}\Big\rbrack
+E^{nr}_{27}\label{eq:4.39}
\ea
\end{prop}
\smallskip

In our analysis of the large time behavior ($t\to\pm\infty$), we shall
require an upper bound of the error terms $E_{2j}^{nr}$ given in:

\begin{prop}
\ba
&&|E^{nr}_{2j}\ |\le\ C_\varphi\ 
 |\lambda|^2\ |A|^2\Big\lbrace\ |A|^2 |A'|\  +\ 
\|\la x\ra^{-\sigma} (B-\zeta_j)^{-1}\  e^{iBt}\  \Pc \varphi^3 \|_2\nn\\ 
&+&\ \|\la x\ra^{-\sigma} (B-\zeta_j)^{-2}\  
\ \i e^{iB(t-s)}\ {\cal O}\left((|A|^3)''\right)\ ds\ 
 \Pc\varphi^3 \|_2\ \Big\rbrace
,\label{eq:4.40}
\ea
where $\zeta_1=\zeta_6=\Omega,\ \zeta_2=\zeta_5=-\Omega,\ \zeta_3=\zeta_4=-3\Omega,$ and
$\zeta_7=3\Omega+i0$.
\end{prop}
In section 7, we shall estimate this expression using the decay
estimates of section 2, in particular Proposition 2.2.
\medskip

The desired form of the $A$-equation is now emerging.
Use of Propositions 4.4 and 4.6 in (\ref{eq:Aeqn}) yields:

\begin{prop}
The amplitude $A(t)$ satisfies the
equation (see Proposition 4.6 for the definition of $\rho(\zeta)$):

 \ba
A'&=&\ {-i\lambda\over 2\Omega}\ \|\varphi\|_4^4\ (\ 3|A|^2A\ +\ A^3 e^{2i\Omega t}\  +\ 
 3|A|^2\ov \ 
e^{-2i\Omega t}\  +\ \ov^3 e^{-4i\Omega t}\ )\nn\\
&&\ -{3\over 4}\ {\lambda^2\over\Omega}\ \Gamma|A|^4A\ -\ {3i\over 4\Omega}\ 
  \lambda^2\ \Big\lbrack
\Lambda\ -\ 5\rho(\Omega)\ +\ 3\rho(-\Omega)\ +\ \rho(-3\Omega)\ \Big\rbrack |A|^4A\nn\\
&&\ -{3\lambda^2\over 4\Omega} A^5\  e^{4i\Omega t}\ \Big\lbrack
i\Lambda\ +\ \Gamma\ -\ \rho(-3\Omega)\ \Big\rbrack\nn\\
&&\ -{3i\lambda^2\over 4\Omega}\ |A|^2A^3\  e^{2i\Omega t}\ \Big\lbrack
-2\ \Lambda\ +\ 2i\Gamma\ +\  \rho(\Omega)\ +\ i\rho(-3\Omega)\ +\ 
 3\rho(-\Omega)\ 
\Big\rbrack\nn\\
&&\ -{3i\lambda^2\over 4\Omega}\  |A|^4\ov\  e^{-2i\Omega t}\  \Big\lbrack
\ 7\rho(\Omega)\ +\ 9\rho(-\Omega)\ +\ \rho(-3\Omega)\ +\ \rho(3\Omega+i0)\ 
 \Big\rbrack\nn\\
&&\ -{3i\lambda^2\over 4\Omega} |A|^2\ov^3\ e^{-4i\Omega t}\ \Big\lbrack
\ 3\rho(-\Omega)\ -\ 2i\rho(-3\Omega)\ +\ \rho(3\Omega+i0)\ +\ 3\rho(\Omega)\ 
 \Big\rbrack\nn\\
&&\ -{3i\lambda^2\over 4\Omega}\ \ov^5\  e^{-6i\Omega
t}\ \Big\lbrack\ i\rho(-3\Omega)\ -\ {4\over3}i\rho(3\Omega+i0)\ \Big\rbrack\ +\ {\bf E}\nn\\
 \label{eq:4.41}
\ea
where
\ba
{\bf E} &=&\ (2i\Omega)^{-1} e^{i\Omega t}\  3\lambda a\int\varphi\eta^2\nn\\
&+&\ (2i\Omega)^{-1}\lambda e^{-i\Omega t}\int\varphi \eta^3\nn\\
&+&\ \sum^7_{j=1} E^{nr}_{2j}\nn\\
&+&\ (2i\Omega)^{-1} e^{-i\Omega t}\Big\lbrack F_2(a,\eta_1)+F_2(a,\eta_3)\Big\rbrack\nn\\
&+&\ F_2(a,\eta^{nr1}_* +\eta_*^{nr2})\label{eq:4.42}
\ea
\end{prop}
Here, $F_2(a,\eta_j)$ is given by (\ref{eq:F1234}). 

In the next section we show how to rewrite 
(\ref{eq:4.41}) in a manner which makes explicit which terms 
determine the large time behavior of the amplitude and phase of $A(t)$.

\bigskip 
\section{Dispersive Hamiltonian normal form}

To analyze the asymptotic behavior of $A(t)$ (or equivalently $a(t)$)
and $\eta(t,x)$ as
 $t\rightarrow\infty$ it is useful to use the idea of
 normal forms \cite{kn:KAM}, \cite{kn:GH}, \cite{kn:SV} from dynamical
systems theory. We derive a perturbed normal form which makes 
 the anticipated large time behavior of solutions transparent.  
\smallskip

\begin{prop}
There exists a smooth near-identity change of variables, $A\mapsto \tilde A$
with the following properties: 
\ba 
\tilde A\ &=&\  A\ +\ h(A,t)\nn\\
 h( A,t)\ &=&\ O(|A|^3),\quad |A|\to0\nn\\
   h(A,t)\ &=&\ h(A,t+2\pi\Omega^{-1}),\label{eq:cov}
\ea
 and such that in terms of $\tilde A$ equation (\ref{eq:4.41}) becomes: 
\ba
{\tilde A}'&=&\ i\lambda c_{21} |\tilde A|^2\tilde A +\lambda^2
d_{32}|\tilde
A|^4\tilde A+i\lambda^2 c_{32}|\tilde A|^4\tilde A\nn\\
&+&\ O(|\tilde A|^7)+{\bf \tilde E}\ ,
\label{eq:tildeAeqn}
\ea
where $\lambda^2 d_{32} =-{3\over4}{\lambda^2\over\Omega} \Gamma<0$.
 The constants $c_{21}$ and $c_{32}$ are real
numbers, explicitly calculable in terms of the coefficients appearing in
(\ref{eq:4.41}). The remainder term $|\tilde{\bf E}|$ is estimable in
terms of
$|{\bf E}|$ for $|\tilde A|<1$ (equivalently
$|A|<1)$.
\end{prop}
\nit{\bf Remarks:} 

\nit The point of this proposition is that in the new
variables the dynamics evidently have a dissipative aspect. Specifically,
neglecting the perturbation to the normal form one has:
\be
\D_t\ |\tilde A|^2\ =\ -{3\over8}\ {\lambda^2\over\Omega}\ \Gamma\ 
                     |\tilde A|^6\ <\ 0
\nn\ee
We therefore refer to  
 (\ref{eq:tildeAeqn}) as a {\it dispersive Hamiltonian
normal form}. In finite dimensional Hamiltonian systems, the normal form
associated with a one degree of freedom Hamiltonian system is an
equation like 
 (\ref{eq:tildeAeqn}), but with all the coefficients of the terms
$|A|^{2m}A$ being purely imaginary. Here we find that resonant coupling to
an infinite dimensional dispersive wave field can lead to a normal form
with general complex coefficients, which in our context implies the  
internal damping effect described above. See section 8 for further
discussion. 
\bigskip

We now present an elementary derivation of the change of variables
(\ref{eq:cov}) leading to (\ref{eq:tildeAeqn}). 
 Equation (\ref{eq:4.41}) is of
the form:
\be
A'\ =\ \sum_{j\in\{3,5\}}\ \sum_{k+l=j}\ \alpha_{kl}\ A^k\ 
 {\overline A}^l\ e^{i(k-l-1)\Omega t}\ +\ {\bf E},
\label{eq:prepnf1}
\ee
where the coefficients $\alpha_{kl}$ can be read off (\ref{eq:4.41}).
The proof we present is quite general and shows, in particular, that 
 the normal
  form for equations like (\ref{eq:prepnf1}) is (\ref{eq:tildeAeqn})
\begin{prop}
 There is a change of variables, as in (\ref{eq:cov}), such that
 equation (\ref{eq:prepnf1}) is mapped to:
\ba
\tilde A'\ &=&\ k_{21} |\tilde A|^2\tilde A 
 + k_{32}|\tilde A|^4\tilde A\nn\\
 &+&\ {\cal O}( |\tilde A|^7)+{\bf \tilde E}\ ,\  {\rm where} 
\label{eq:generaltildeAeqn}
\ea
\ba
{\bf \tilde E}\ &=&\ {\bf E}\circ (I+h),\nn\\
k_{21}\ &=&\ \alpha_{21}\nn\\
k_{32}\ &=&\  \alpha_{32}\ +\ (2i\Omega)^{-1}\left[\ 
{3\over2}|\alpha_{03}|^2\ -\ 2\alpha_{30}\alpha_{12}\ +\
 2 |\alpha_{12}|^2\ \right]\label{eq:kij}
\ea
\end{prop}
 The conclusion in Proposition 5.1 concerning the "damping 
  coefficient" $d_{32}$
 depends on the particular properties of the coefficients in
 (\ref{eq:Aeqn}).  In particular we have from (\ref{eq:Aeqn}) and
 (\ref{eq:prepnf1}) that:
 \ba
 \alpha_{21}\ &=&\ \alpha_{12}\ =\ 
  {3\lambda\over 2i\Omega}\|\varphi\|_4^4,\nn\\
 \alpha_{03}\ &=&\ {\lambda\over 2i\Omega}\|\varphi\|_4^4,\nn\\
 \alpha_{32}\ &=&\ -{3\lambda^2\over4\Omega}\Gamma\ -\ 
	  {3i\lambda^2\over 4\Omega}\ \left[ \Lambda\ -\ 5\rho(\Omega)\ +
	   \ 3\rho(-\Omega)\ +\ \rho(-3\Omega)\ \right].
\label{eq:alphaij}
\ea
From these formulae, we have $\lambda^2 d_{32}$ is given by the real part of
$\alpha_{32}$, the "singular" Fermi golden rule contribution.
 
\nit{\bf Remark:}
The construction of the map $A\mapsto {\tilde A}$ can
 be applied as well to equations of the form (\ref{eq:prepnf1}) where
 the right hand side is an 
  arbitrary expansion in powers of $A$ and ${\overline A}$. 
\medskip

To make the structure clear we write out the equation with
a particular ordering of terms:
\ba
A'\  &=&\ \alpha_{21}\ |A|^2A\ +\ \alpha_{32}\ |A|^4A\ +\
  O_3(A)\ +\ O_5(A)\ +\  {\bf E},
\label{eq:prenfm2}\\
\quad&&{\rm\ where\ }\nn\\
\ O_3(A)\ &=&\ \alpha_{30}\ A^3\ e^{2i\Omega t}\ +\ \alpha_{12}\
  A{\overline A}^2\ e^{-2i\Omega t}\ +\ \alpha_{03}\ {\overline A}^3\
 e^{-4i\Omega t},\quad {\rm and\ }\label{eq:O3}\\
\ O_5(A)\ &=&\ \alpha_{50}\ A^5\ e^{4i\Omega t}\ +\
\alpha_{41}\ A^4 {\overline A} \ e^{2i\Omega t}\ +\
\alpha_{23}\ A^2 {\overline A}^3 \ e^{-2i\Omega t}\nn\\
\quad &+&\ \alpha_{14}\ A {\overline A}^4 \ e^{-4i\Omega t}\ +\
 \alpha_{05}\ {\overline A}^5 \ e^{-6i\Omega t}\
 \label{eq:O5}
\ea
Note that each term in $O_3(A)$ and $O_5(A)$ is of the 
form: oscillatory function of $t$ times order ${\cal O}(|A|^3)$ 
 or ${\cal O}(|A|^5)$.

The computations that follow, though elementary, are rather lengthy so
we first outline the strategy of our proof.
Integration of (\ref{eq:prenfm2}) gives
\be
A\ =\ A_0\ +\ \int_0^t\ \alpha_{21}\ |A|^2A\ +\ \alpha_{32}\ |A|^4A\ +\
  O_3(A)\ +\ O_5(A)\ +\  {\bf E}\ ds.
\label{eq:Aint}
\ee
The idea is that terms with explicit periodic oscillations average to
zero and can be neglected in
 determining the large time behavior of the solution. 
Our strategy is now to expand the explicitly oscillatory terms using 
repeated  
 integrations by parts and to make explicit all terms up to and including order
$|A|^5$. The computation has two stages. In stage one, after repeated
integration by parts the equation (\ref{eq:Aint}) is expressed in the
equivalent form:
\ba
A(t)\ -\ H_1(A(t),t)\ &=&\ A_0\ -\ H_1(A_0,0)\nn\\
&+&\ \int_0^t\ {\rm \ "resonant"\ terms\ like\ }\ |A|^2A\ ,\ |A|^4A\
ds\nn\\
&+&\ \int_0^t\ {\rm terms\ of\ type }\ O_5(A)\ ds\ +\ {\rm higher\ order\ corrections}
\label{eq:Aint1}
\ea
This suggests the change of variables $A\mapsto A_1=A(t)\ -\
H_1(A(t),t)$, giving
\ba
A_1(t)\  &=&\ A_{10}\ 
 +\ \int_0^t\ {\rm terms\ like\ }\ |A_1|^2A_1\ ,\ |A_1|^4A_1\
ds\nn\\
&+&\ {\rm terms\ of\ type }\ O_5(A_1)\ +\ {\rm higher\ order\ corrections}.
\label{eq:Aint2}
\ea
The latter equation is equivalent to a differential equation
of the form:
\ba
A_1'(t)\ &=&\ \ {\rm terms\ like\ }\ |A_1|^2A_1\ ,\
|A_1|^4A_1\nn\\
&+&\ {\rm terms\ of\ type\ } O_5(A_1)\ +\ {\rm higher\ order\
corrections}.
\label{eq:Aint1diff}
\ea
which is a step closer to  the form of the equation  we seek. A second
iteration of this process yields the result.
\medskip

We now  embark on the details of the proof.
\medskip

\nit{\it Expansion of} $\int_0^t\ O_3(A)\ ds$:
\medskip

Integration of the expression in (\ref{eq:O3}) gives
\ba
\int_0^t\  O_3(A)\ ds\ &=&\ h_3(A(t),t)\ -\ h_3(A_0,0)\nn\\ 
&-&\ {3\alpha_{30}\over 2i\Omega}\ \int_0^t\ e^{2i\Omega s}A^2\ A'\ ds\ 
 +\ {\alpha_{12}\over 2i\Omega}\ \int_0^t\ e^{-2i\Omega s}\ \left( A\ {\overline
 A}^2\right) '\ ds\nn\\
&+&\ {\alpha_{03}\over 4i\Omega}\ \int_0^t\ e^{-4i\Omega s} \left( {\overline A}^3 \right)'
\ ds,
\label{eq:intO3-1}
\ea
where
\be
h_3(A,t)\ =\ {\alpha_{30}\over 2i\Omega}\ A^3\ e^{2i\Omega t}
\  -\ {\alpha_{12}\over 2i\Omega}\ A\ {\overline A}^2\ e^{-2i\Omega t}
\  -\ {\alpha_{03}\over 4i\Omega}\ {\overline A}^3\ e^{-4i\Omega t}
\label{eq:h3}
\ee

We now replace $A'$ in (\ref{eq:intO3-1}) by 
 its abbreviated expression given in 
 (\ref{eq:prenfm2}).
Thus we have:
\ba
\int_0^t\  O_3(A)\ ds\ &=&\ h_3(A(t),t)\ -\ h_3(A_0,0)\nn\\ 
 &-& {3\alpha_{30}\over 2i\Omega}\ \int_0^t\ e^{2i\Omega s}A^2\ 
\left[ \alpha_{21}\ |A|^2A\ +\  
  O_3(A)\ +\ {\cal O}(|A|^5)\ +\  {\bf E}\ \right]\ ds\nn\\
 &+& {\alpha_{12}\over 2i\Omega}\ \int_0^t\ e^{-2i\Omega s}{\overline A}^2\
 \left[ \alpha_{21}\ |A|^2 A\ +\  O_3(A)\ +\   
  {\cal O}(|A|^5)\ +\ {\bf E}\ \right]\ ds\nn\\
&+& {\alpha_{12}\over 2i\Omega}\ \int_0^t\ e^{-2i\Omega s}
\ 2|A|^2\
 \left[ {\overline \alpha}_{21}\ |A|^2{\overline A}\ +\ {\overline O}_3(A)\    
  +\  {\cal O}(|A|^5)\ +\ \overline{{\bf E}}\ \right]\ ds\nn\\
 &+& {3\alpha_{03}\over 4i\Omega}\ \int_0^t\ e^{-4i\Omega s}{\overline
A}^2\
 \left[ {\overline \alpha}_{21}\ |A|^2{\overline A}\ +\ {\overline O}_3(A)\ 
+\  {\cal O}(|A|^5)\ +\ \overline{\bf E}\ \right]\ ds  
\label{eq:intO3}
\ea

Substitution of (\ref{eq:O3}) into (\ref{eq:intO3}) and integrating by
parts, we arrive at the  
following expression:
\ba
\int_0^t\ O_3(A)\ ds\ &=&\ h_3(A(t),t)\ +\ h_{3a}(A(t),t)\ -
 h_3(A_0,0)\ -\ h_{3a}(A_0,0)\nn\\
&+&\  {1\over 2i\Omega}\ \left(  {3\over2}\
|\alpha_{03}|^2\ -\ 2\alpha_{30}\alpha_{12}\ +\ 2|\alpha_{12}|^2
 \right)\ \int_0^t\ |A|^4A\
ds\nn\\
&+&\ \int_0^t\ {\cal O}\left( |A|^2(|A|^5+|{\bf E}|)\right)\ ds,
\label{eq:intO3a}
\ea
where
\ba
h_{3a}(A,t)\ &=&\ \left[\ -{\alpha_{12}{\overline \alpha}_{03}\over2\Omega^2}\
+\ {3\alpha_{30}\alpha_{21}\over 4\Omega^2}\ \right] 
              A^4{\overline A}\ e^{2i\Omega t}\ -\ 
{3\alpha_{30}^2\over 8\Omega^2}\
A^5\ e^{4i\Omega t}\nn\\
&+&\ {1\over 4\Omega^2}\  \left(\ \alpha_{21}\alpha_{12}\ +\
{3\over2}\alpha_{03}{\overline \alpha}_{12}\ -\ 3\alpha_{30}\alpha_{03}\
 +\ 2\alpha_{12}\overline\alpha_{21}\right)\ 
A^2{\overline A}^3\ e^{-2i\Omega t}\nn\\ 
&+&\ {1\over 8\Omega^2}\  \left(\ \alpha_{12}^2\ +\
{3\over2}\alpha_{03}{\overline \alpha}_{21}\ +\
2\alpha_{12}\overline\alpha_{30}\ \right)\  
\ A{\overline A}^4\ e^{-4i\Omega t}\nn\\
&+&\ {1\over 4\Omega^2}\  \left(\ {1\over3}\alpha_{03}\alpha_{12}\ +\
{1\over2}\alpha_{03}{\overline \alpha}_{30}\ \right)\  
{\overline A}^5\ e^{-6i\Omega t}.  
\label{eq:h3a}
\ea
\medskip
Referring back to (\ref{eq:prenfm2}), we see we must now obtain an
\medskip

\nit{\it Expansion of} $\int_0^t\ O_5(A)\ ds$:
\medskip

From (\ref{eq:O5}) we have, after integration by parts:
\ba
\int_0^t\ O_5(A)\ ds\ &=& \ h_{3b}(A(t),t)\ -\ h_{3b}(A_0,0)\nn\\
 &+&\ \int_0^t\ {\cal O}\left(\ |A|^4(|A|^3\ +\ |{\bf E}|)\ 
\right)\ ds,
 \label{eq:intO5}
\ea
where
\ba
h_{3b}(A,t)\ &=&\ {\alpha_{50}\over 4i\Omega}\ A^5\ e^{4i\Omega t}\ +\ 
 {\alpha_{41}\over 2i\Omega}\ A^4 {\overline A}\ e^{2i\Omega t}\ -\ 
 {\alpha_{23}\over 2i\Omega}\ A^2\ {\overline A}^3\ e^{-2i\Omega t}\nn\\
&-&\ {\alpha_{14}\over 4i\Omega}\ A {\overline A}^4\ e^{-4i\Omega t}\ -\
 {\alpha_{05}\over 6i\Omega}\ {\overline A}^5\ e^{-6i\Omega t}.
\label{eq:h5}
\ea
Therefore from (\ref{eq:Aint}) and our computations we have:
\ba
&&\ A\ -\ h_3(A,t)\ -\ h_{3a}(A,t)\ -\ h_{3b}(A,t)\ \nn\\
&&\ =\ A_0\  -\ h_3(A_0,0)\ -\ h_{3a}(A_0,0)\ -\ h_{3b}(A_0,0)\nn\\ 
&+& \int_0^t\ \alpha_{21}\ |A|^2A\ +\ \alpha_{32}\ |A|^4A\
ds\nn\\
&+&\ {1\over 2i\Omega}\ \left(\ 
 {3\over2}|\alpha_{03}|^2\ -\ 2\alpha_{30}\alpha_{12}\ +\
2|\alpha_{12}|^2\  \right)\ \int_0^t\
|A|^4A\ ds\ +\ \int_0^t\ {\bf E}\ ds\nn\\
&+&\ \int_0^t\ {\cal O}\left(\ |A|^4 (|A|^3\ + |{\bf E}|)\ \right)\ 
 +\ {\cal O}\left(\ |A|^2 (|A|^5\ + |{\bf E}|)\ \right)\ ds.
\label{eq:pre-cov} 
\ea

This suggests the change of variables:
\ba
A_1\ &\equiv&\ A\ -\ H_1(A,t),\quad {\rm where}\nn\\
H_1(A,t)\ &=&\ h_3(A,t)\ +\ h_{3a}(A,t)\ +\ h_{3b}(A,t),
\label{eq:A1}
\ea
which for small $|A|$, is a near-identity change of variables. 
Using this change of variables we have that  (\ref{eq:pre-cov}) becomes
\ba
A_1\ &=&\ A_{10}\ +\ \int_0^t\ \alpha_{21}\ |A_1|^2A_1\ +\ \alpha_{32}\ |A_1|^4A_1\
ds\nn\\
&+&\ {1\over 2i\Omega}\ \left(\  
{3\over2}|\alpha_{03}|^2\ -\ 2\alpha_{30}\alpha_{12}\ +\ 
 2|\alpha_{12}|^2\ \right)\ \int_0^t\
|A_1|^4A_1\ ds\nn\\
 &+&\ \int_0^t\ \left(\ 2\alpha_{21}\ |A_1|^2\ H_1(A_1,s)\ + A_1^2\
{\overline H_1}(A_1,s)\ \right)\ ds\   
 +\ \int_0^t\ {\bf E_1}\ ds\nn\\
&+&\ \int_0^t\ {\cal O}\left(\ |A_1|^4 (|A_1|^3\ + |{\bf E_1}|)\ \right) \ 
 +\ {\cal O}\left(\ |A_1|^2 (|A_1|^5\ + |{\bf E_1}|) \right)\ ds.
\label{eq:A1eqn}
\ea

We expand the terms involving $H_1(A,t)$. Using integration by parts, as
above, we obtain:
\be
2\alpha_{21}\int_0^t\ |A_1|^2\ H_1\ ds\ = h_{5a}(A_1,t)\ -\ h_{5a}(A_{10},0)\
+\ \int_0^t\ {\cal O}\left(\ |A_1|^4 (|A_1|^3\ + |{\bf E_1}|)\ \right)\
ds,
\label{eq:intH1a}
\ee
and 
\be
\alpha_{21}\int_0^t\ A_1^2\ {\overline H}_1\ ds\ = h_{5b}(A_1,t)\ -\ 
 h_{5b}
(A_{10},0)\
+\ \int_0^t\ {\cal O}\left(\ |A_1|^4 (|A_1|^3\ + |{\bf E_1}|)\ \right)\
ds,
\label{eq:intH1b}
\ee
where
\ba
h_{5a}(A,t)\ &=&\ -{\alpha_{21}\alpha_{30}\over2\Omega^2}\ 
e^{2i\Omega t}\ A_1^4\ {\overline A_1}\ -\
{\alpha_{21}\alpha_{12}\over 2\Omega^2}\ 
e^{-2i\Omega t}\ A_1^2\ {\overline A_1}^3\nn\\
&-&\ {\alpha_{21}\alpha_{03}\over 8\Omega^2}\ 
e^{-4i\Omega t}\ A_1\ {\overline A_1}^4 \label{eq:h5a}\\
h_{5b}(A,t)\ &=&\ 
-{\alpha_{21}{\overline \alpha}_{30}\over 4\Omega^2}\ 
e^{-2i\Omega t}\ A_1^2\ {\overline A_1}^3\ -\ 
 {\alpha_{21}{\overline \alpha}_{12}\over 4\Omega^2}\ 
e^{2i\Omega t}\ A_1^4\ {\overline A_1}\nn\\
&-&\ {\alpha_{21}{\overline \alpha}_{30}\over 16\Omega^2}\ 
e^{4i\Omega t}\ A_1^5 
 \nn\\
H_{5}(A,t)\  &=&\ h_{5a}(A,t)\ +\ h_{5b}(A,t)
\label{eq:H5ab}
\ea
Thus
\ba
A_1\ -\ H_5(A_1(t),t)\ &=&\ A_{10}\ -\ H_5(A_{10},0)\nn\\
 &+&\  \int_0^t\ \alpha_{21}\ |A_1|^2A_1\ +\ \alpha_{32}\ |A_1|^4A_1\
ds\nn\\
&+&\ {1\over 2i\Omega}\ \left(\ 
{3\over2}|\alpha_{03}|^2\ -\ 2\alpha_{30}\alpha_{12}\ +\ 
 2|\alpha_{12}|^2\ \right)\ \int_0^t\
|A_1|^4A_1\ ds\nn\\
&+&\ \int_0^t\ {\cal O}\left(\ |A_1|^7\ +\ |A_1|^2 |{\bf E_1}|\ +\
 |{\bf E_1}| \right)\ ds.
 \nn\\
\label{eq:preA2eqn}
\ea
Now define
\be
\tilde A\ \equiv\ A_1\ -\ H_5(A_1,t).
\label{eq:tAdef}
\ee
In terms of this new variable we have:
\ba
\tilde A\ &=&\ \tilde A_0\ +\ \int_0^t\ \alpha_{21}\ |\tilde A|^2\tilde
 A\ +\
\alpha_{32}\ |\tilde A|^4\tilde A\
ds\nn\\
&+&\ {1\over 2i\Omega}\ \left(\ 
{3\over2}|\alpha_{03}|^2\ -\ 2\alpha_{30}\alpha_{12}\ +\ 
 2|\alpha_{12}|^2\  \right)\ \int_0^t\
|\tilde A|^4\tilde A\ ds\nn\\
&+&\ \int_0^t\ {\cal O}\left(\ |\tilde A|^7\ +\ |{\bf \tilde
 E}|\ \right)\ ds, 
\label{eq:tAeqn}
\ea
where $|{\bf \tilde E}|$ is estimable in terms of $|{\bf E}|$ for
$|A|<1$.
This completes the proof.

%%%%%%%%%%%%%%%%%%%%%%%%
\bigskip
\section{Large $t$ behavior of solutions to the perturbed normal form
equations}
\medskip

We now consider the large time behavior of solutions to the
ordinary differential equations of the form (\ref{eq:tildeAeqn}).
In particular, we compare the behavior of solutions 
 of (\ref{eq:tildeAeqn}) 
to those of the equation with $\tilde{\bf E}$ set equal to zero.

Thus  we consider the equations: 
\ba
\beta'\ &=&\  ic_{21}|\beta|^2\beta\ 
 +\ (ic_{32}-\gamma)\ |\beta |^4\beta\ 
 +\ {\bf
Q}(t),
\label{eq:pnfm}\\
\alpha'\ &=&\ ic_{21}|\alpha|^2\alpha\ +\ (ic_{32}-\gamma)\ 
 |\alpha|^4\alpha,
\label{eq:nfm}
\ea
where $\gamma>0$.
We first consider the unperturbed equation, (\ref{eq:nfm}). 
Multiplication of (\ref{eq:nfm}) by $\overline\alpha$ 
 and taking the real part of
the resulting equation yields the equation:
\be 
r'\ =\ -2\ \gamma\ r^3,\qquad r\ =\ |\alpha|^2.
\label{eq:amp2}
\ee
Integration of (\ref{eq:amp2}) yields:
\be
r^2(t)\ =\ r_0^2\ \left( 1\ +\ 4\ \gamma\ r_0^2\ t \right)^{-1}.
\nn\ee

To prove the above lemma, we begin by multiplying  (\ref{eq:pnfm}) by 
 $\overline \beta$ and taking the real part of the resulting equation. This
gives:
\be
r'(t)\ =\ -2\ \gamma\ r^3(t)\ +\ {\bf Q}(t)\overline \beta(t)\ +\ \overline 
{\bf Q}(t)\
\beta(t),\nn
\ee
which implies the differential inequality:
\be
r'(t)\ \le\ -2\ \gamma\ r^3(t)\ +\ 2\ |{\bf Q}(t)|\ r^{1\over2}(t).
\label{eq:ineq}
\ee

We now prove 
\begin{lem}
Suppose $r(t)\ =\ |\beta(t)|^2$ satisfies (\ref{eq:ineq}) with:
\be
|{\bf Q}(t)|\ \le\  Q_0\ \la t\ra^{-{5\over4}-\delta},\quad \delta\ge0.
\label{eq:Qbound}
\ee
Then, 
\be
|\beta(t)|^4\ \le\ \left( 1+4|\beta_0|^4\gamma t\right)^{-1}\ 
  \left(\ 2|\beta_0|^4\ +\ {\ m_*^2\ Q_0^{8\over5}\over
 \gamma^{3\over5}\ \la t\ra^{{8\over3}\delta} }\ \right), 
\label{eq:Aest}
\ee
where $m_*\ =\ \max\{1,4\gamma |\beta_0|^4\}$.
\end{lem}
\nit{\it proof:}
 Use of (\ref{eq:Qbound}) in (\ref{eq:ineq}) gives the inequality
\be
r'(t)\ \le\ -2\gamma r^3(t)\ +\ 2Q_0\ \la t\ra^{-{5\over4}-\delta}\
r^{1\over2}(t).\nn
\ee 
Multiplication by $r(t)$ gives:
\be
z'(t)\ \le -4\gamma\ z^2(t)\ +\ 4Q_0\ \la t\ra^{-{5\over4}-\delta}\
z^{3\over4}(t),\quad {\rm where}\ z(t)\ =\ r^2(t).
\label{eq:zeqn}\ee
Note that the equation 
\be \zeta'(t)\ =\ -4\gamma\ \zeta^2(t), \quad  \zeta(0)\ =\ z_0\ =\
|\beta_0|^4\nn\ee
has solutions:
\be \zeta(t)\ =\ \ z_0\ \left( 1\ +\ 4\ z_0\gamma t\right)^{-1}.
 \nn\ee
Anticipating this as the dominant behavior for large $t$, we define:
\be z(t)\ \equiv\ \zeta(t)\ R(t).\label{eq:zR}\ee
Substitution into (\ref{eq:zeqn}) and simplifying  gives:
\be
R'(t)\ \le\ 
{-4\gamma z_0\over 1\ +\ 4\ z_0\gamma t}\ R\ (R-1)\ +\
 C\ {Q_0\over |z_0|^{1\over4}}\ \left( 1 + 4z_0\gamma t\right)^{1\over4}
\la t\ra^{-{5\over4}-\delta}\ R^{3\over4}(t).\label{eq:Reqn1}
\ee
We now consider the last term in (\ref{eq:Reqn1}). Noting that
\be
(1+t)^{-1}\le m_*\ (1+4\gamma z_0 t)^{-1},\quad 
 m_*\ =\ \max\{1,4\gamma z_0\},  
\nn\ee
 we have for any $\veps > 0$:
\ba
C\ {Q_0\over |z_0|^{1\over4}}\ { \left( 1 + 4z_0\gamma
t\right)^{1\over4}\over 
\la t\ra^{{5\over4}+\delta} }\ R^{3\over4}(t)\ &\le&\ 
    C\ {Q_0\over |z_0|^{1\over4}}\ { m_*^{1\over4}\over \la
   t\ra^{1+\delta} } \ R^{3\over4}\nn\\
&\le&\ C\ {Q_0\ m_*^{5\over8} \over \varepsilon\ |z_0|^{1\over4}
                                   \la t\ra^{{5\over8} + \delta}}\ 
\times  {\varepsilon\ R^{3\over4}(t)\over (1 + 4z_0\gamma t)^{3\over8}}\nn\\
&\le&\ C_1\ { Q_0^{8\over5}\ m_*^{3\over5} \over \varepsilon^{8\over5}\
z_0^{2\over5}\ \la t\ra^{1+{8\over5}\delta} }\ +\ 
C_2\ { \varepsilon^{8\over3}\ R^2(t) \over (1 + 4z_0\gamma t)}.
\nn
\ea
The last inequality follows from the inequality: $ab\le p^{-1}(\veps a)^p+q^{-1}(b/\veps)^q,\quad p^{-1} + q^{-1}=1$, for the choice $p={8\over3}$ and
$q={8\over5}$.

This last estimate can now be used in (\ref{eq:Reqn1}) and implies:
\ba
R'\ &\le&\ {-4\gamma z_0 + C_2 \veps^{8\over3}\over 1 + 4z_0\gamma t}\
R^2(t)\ +\ {4\gamma z_0\over 1 + 4z_0\gamma t}\  R(t)\ +\ 
 C_1\  { Q_0^{8\over5}\ m_*^{3\over5} \over \varepsilon^{8\over5}\ 
z_0^{2\over5}\ \la t\ra^{1+{8\over5}\delta} }\nn\\
&& {\rm \ which\  when\  setting}\ C_2\veps^{8\over3}=2\gamma z_0,\ {\rm
\ is\  }\nn\\
&\le&\ -{2\gamma z_0 \over 1 + 4z_0\gamma t}\ R\ (R-2)\ +\ 
 C {Q_0^{8\over5}\ m_*\over \gamma^{3\over5}z_0 \la
t\ra^{1+{8\over3}\delta}}.
\label{eq:Reqn3}
\ea

Now set $R\ =\ 2\ +\ S$. Since the term proportional to $S^2$ is
negative, we obtain
\be
S'\ \le\ {-4\gamma z_0\over 1\ +\ 4\ z_0\gamma t}\ S \ +\ 
 C {Q_0^{8\over5}\ m_*\over \gamma^{3\over5}z_0 \la
t\ra^{1+{8\over5}\delta}}.
\label{eq:Seqn1}\ee

Multiplication by $1 + 4\ z_0\gamma t$ yields:
\ba
\D_t\left[\ (1\ +\ 4\ z_0\gamma t)\ S(t)\ \right]
\ &\le&\  C {Q_0^{8\over5}\ m_*\over \gamma^{3\over5}z_0}\ 
{1 + 4\ z_0\gamma t\over (1 + t)}\ {1\over \la t\ra^{{8\over5}\delta}}\nn\\
&\le&\ C {Q_0^{8\over5}\ m_*^2\over \gamma^{3\over5}z_0}\ 
{ 1\over \la t\ra^{{8\over5}\delta} }.
\label{eq:Seqn2}
\ea
Integration of (\ref{eq:Seqn2}) from $0$ to  $t$ implies:

\be
S(t)\ \le\ C\ {Q_0^{8\over5}\ m_*^2\over \gamma^{3\over5}z_0}\ 
{1\over \la t\ra^{{8\over5}\delta}}. 
\label{eq:Sest}
\ee
The Lemma now follows from (\ref{eq:Sest}) and the relation:
\be
|\beta(t)|^4\ \equiv z(t)\ \equiv \zeta(t)\ R(t)\ \equiv\ \zeta(t)\ 
\left(\ 2\ +\ S(t)\ \right).
\nn
\ee

\bigskip
\bigskip
\section{ Asymptotic behavior of solutions to the nonlinear Klein-
Gordon equation}

In \S2 we proved that local in time solutions, $\bu(t)$, 
 exist in the space  $C^0(-T^*,T_*;\bXz)$, for some $T_*, T^* >0$. 
We now study obtain the required {\it 'a priori} bounds to ensure
 (a) persistence of the solution, $\bu(t)$, as a continuous function with values
in  $\bXz$  ($T^*=T_*=\infty$) and 
 (b)  the decay of solutions as $t\to\pm\infty$ in
suitable norms. Due to the linear estimates
of section 2, we require more stringent hypotheses on
$u_0$ and $u_1$. These linear estimates require finiteness of  $W^{1,4/3}$ and
$W^{1,8/7}$ norms which, by interpolation, are controlled  under the assumption
 $\bu_0\in\bX$ (see
section 2). Specifically, $u_0\in W^{2,2}\cap W^{2,1}$ and $u_1\in
W^{1,2}\cap W^{1,1}.$  

Using the results of the previous section, the original dynamical
systems (\ref{eq:nlkg1}) can now be rewritten as:
\ba 
u(x,t) &=&  a(t)\ \varphi(x) + \eta(x,t),\nn \\
\eta(x,t) &=& \eta_1(x,t) + \eta_2(x,t) + \eta_3(x,t),\nn \\
a(t) &=& A(t)\ e^{i\Omega t} + {\bar A}(t)\ e^{-i\Omega t},\nn \\
A(t) &=& \tA(t) + h(\tA(t),t),\label{eq:decomposition}
\ea
Here, $h(\tA,t)$ is a smooth periodic function of $t$, cubic in $\tA$ for $\tA
\to0$, $\eta_1$ and $\eta_2$ are defined by (\ref{eq:eta1eqn})-
(\ref{eq:eta2eqn}), and  $\tA(t)$ satisfies the perturbed normal form equation:

\be \tA'\ =
 \  i\lambda c_{21}\ |\tA|^2\tA\  -\ \gamma\ |\tA|^4\tA \ +\ 
i\lambda^2c_{32}|\tA|^4\tA\  +\ {\cal O}(|\tA |^7)\ +\  {\tilde
{\bf E}}(t,a,\eta). \label{eq:taeqn}\ee
Here $\tilde {\bf E} = {\bf E}\circ (I+h)$,
  with ${\bf E}$ given by (\ref{eq:4.42}).
\bigskip

In (\ref{eq:taeqn}), 
\be \gamma\ =\ {3\over4}\ {\lambda^2\over\Omega}\ \Gamma,\qquad
\Gamma\ \equiv\ {\pi\over3\Omega}\ \left( \Pc\varphi^3,\delta(B-3\Omega) \Pc\varphi^3\right)>0.
 \label{eq:constants}\ee
The constants $c_{21}$ and $c_{32}$ are real numbers which are
computable by the algorithm presented in section 5.
\medskip

The above definitions of $A$, $\tilde A$, and $h$, together with the
estimates proved below, can be used to verify in a straightforward
manner the assertions of Theorem 1.1 concerning $R(t)\equiv |A(t)|$, 
and $\theta(t)\equiv \arg{A(t)}$.

To proceed with a study of the $t\to\infty$ behavior, recall that:

\ba &&\left(\D_t^2 + B^2\right)\eta_1 = 0,\qquad\qquad 
    \eta_1(x,0)\ = \ \Pc u_0,\ \D_t\eta_1(x,0)=\ \Pc u_1, \label{eq:eta1} \\
    &&\left(\D_t^2 + B^2\right)\eta_2 = \lambda
          a^3\ \Pc\varphi^3,\qquad
    \eta_2(x,0)=0,\ \D_t\eta_2(x,0)=0, \label{eq:eta2} \\
   &&\left(\D_t^2 + B^2\right)\eta_3 = 
      \lambda \Pc\left( 3a^2\varphi^2\eta +3a\varphi\eta^2
                       +\eta^3 \right),\quad
   \eta_3(x,0)\ =\ \D_t\eta_3(x,0)=0. \label{eq:eta3}
\ea
 
 To motivate the strategy, we 
 first  argue heuristically. Since $\gamma > 0$, if $\tilde{\bf E}$
is negligible, then $|\tA|\sim \jt^{-1/4}$ as $t\to\infty$,
and therefore by (\ref{eq:aA}), $a\sim \jt^{-1/4}$, as
$t\to\infty$. It then follows from (\ref{eq:eta2}) that, in  
 appropriate norms, that $\eta_2\sim \jt^{-3/4}$ and 
 $\eta_3\sim \jt^{-1+\delta}$ for some $\delta>0$.

To make all this precise requires {\it \'a priori} estimates on the
system the above equations. 
We now recall Lemma 6.1 concerning equations of the
form (\ref{eq:taeqn}). This result gives conditions ensuring that $\tA$
behaves like the solution of the equation obtained by setting $\tilde{\bf E}$
to zero. 

Our next task is to obtain an upper bound on $\tilde{\bf E}(t)$ 
 in (\ref{eq:4.42}) of the form (\ref{eq:Qbound}). Since the equation for 
$\tA(t)$ is coupled to that of $\eta$, the factor $Q_0$ in
(\ref{eq:Qbound})  
 will depend on $\eta$ and $\tA$. The next proposition, together with
Proposition 4.7, provides estimates for the individual terms in
$\tilde{\bf E}$. 

It is convenient to introduce the notation:
\be 
 [A]_\alpha(T)\ = \ \sup_{0\le t\le T}\jt^\alpha |A(t)|
 \label{eq:norm1}
\ee
\be [ \eta ]_{p,\alpha}(T)\ = \ 
      \sup_{0\le t\le T}\jt^\alpha 
                    ||\eta(t)||_p. \label{eq:norm2} 
\ee
To avoid cumbersome notation, where it should cause no confusion, we
shall abbreviate expressions like $[\cdot\cdot\cdot](T)$ by
$[\cdot\cdot\cdot]$,  until it is necessary to make the dependence on
$T$ explicit. In the estimates below, we shall often use the notation $C_\varphi$
to denote a constant depending on some $W^{k,p}$ norm of the bound state,
$\varphi$. Under our hypotheses, $\varphi$ is a sufficiently smooth and
exponentially decaying function for which these norms are finite; see
\cite{kn:Agmon}.

Because of Proposition 5.1,
  we need only estimate the terms of ${\bf E}$,
given in (\ref{eq:4.42}). 
The following proposition is the main step toward the estimate on 
 ${\bf E}$. 
\bigskip

\begin{prop} {\it Estimates on the terms in ${\bf E}(t)$:}

For $0\le s\le t$, the terms of ${\bf E}(s)$, as defined in 
 (\ref{eq:4.42}), are estimated as follows:
\ba
{\rm (i)}\ \left|\lambda a\int\varphi \eta^2\right| &\le&
 |\lambda|\ C_\varphi\ 
[A]_{1/4}\ [\eta]_{8,3/4}^2\ \jt^{-7/4}, \label{eq:e1}\\
{\rm (ii)}\ \left|\lambda\int\varphi \eta^3\right| &\le&
 |\lambda|\ C_\varphi\ [\eta]_{8,3/4}^3\ \jt^{-9/4}, \label{eq:e2}\\
{\rm (iii)}\ \left| F_2(a,\eta_1)\right|&\le& |\lambda|\ 
[A]_{1/4}^2\ \|\bu_0\|_{\bX}\ \jt^{-13/8}, \label{eq:e3}\\
{\rm (iv)}\ \left| F_2(a,\eta_3)\right| &\le& 
 C_\varphi\ |\lambda|\ [A]_{1/4}^2\ C\left(\ 
 [A]_{1/4},[\eta]_{8,3/4},[B\eta_3]_{4,1/4+\sigma_0},
 \|\bu_0\|_{\bX}\ \right)
 \  \jt^{-5/4-\sigma_0}, \nn\\
 \label{eq:e4}\\
{\rm (v)}\  \left| F_2(a,\eta_*^{nr1})\right|&\le&
 C_\varphi\ |\lambda|^2\ |A_0|^2\ [A]_{1/4}^2\ \jt^{-2},
\label{eq:e5}\\
{\rm (vi)}\ \left| F_2(a,\eta_*^{nr2})\right| &\le&
C_\varphi\ |\lambda|^2\ [A]_{1/4}^2\ [|A|^2|A'|]_{5/4}\  \jt^{-7/4},
\label{eq:e6}\\
{\rm (vii)}\  \left| E_{2j}^{nr}\right| &\le& |\lambda|\ C_\varphi\  
\left([|A|^4|A'|]_{7/4}\ + \ [A]_{1/4}^2\ +\ [|A|^2|(A^3)''|]_{7/4}\right)\ 
\ \jt^{-13/8}\nn\\
 \label{eq:e7}.
\ea
In (iv), $C(r_1,r_2,r_3,r_4)$ is bounded for $\sum_j|r_j|$ bounded and tends to
zero as $\sum_j|r_j|$ tends to zero; see Proposition 7.4.
\end{prop}
\bigskip
\nit We now embark on the proof of this proposition. 
  Parts (i) and (ii) follow by H\"older's inequality. To prove part (iii), apply
  H\"older's inequality and the linear propagator estimates of Theorem 2.3.
We now focus on (iv)-(vi).
\medskip

\centerline{\it Estimation of $ F_2(a, \eta_3)$}

Recall (see \ref{eq:F1234}) that 
\be 
F_2(a, \eta_3) = 3\lambda a^2\int\varphi^3\eta_3.
\nn\ee
We start by estimating $||\eta_3||_8$.
We first express $\eta_3$, defined in (\ref{eq:eta3}), as
\be \eta_3(t) =\lambda\ \sum_{j=1}^3\
\int_{I_j}E_1(t-s)\Pc\left(3a^2\varphi^2\eta\ +\ 3a\varphi\eta^2\ +\
\eta^3\right)\ ds, \label{eq:eta3a}
\ee
where
\be I_1=[0,t/2],\ I_2=[t/2,t-1],\ {\rm and }\ I_3=[t-1,t].\nonumber\ee
We estimate each integral separately.
\be
 ||\eta_3(t)||_8\le |\lambda|\sum_{j=1}^3\ \int_{I_j}
||E_1(t-s)\Pc\left(3a^2\varphi^2\eta\ +\ 3a\varphi\eta^2\ +\
\eta^3\right)||_8\ ds.\nonumber
\ee
The integrands are estimated as follows using the linear estimates
 of Corollary 2.1:
\ba
\int_{I_1}||E_1(t-s)\Pc\{ \cdot\cdot\cdot\}||_8\ ds &\le& 
      C\int_0^{t/2} |t-s|^{-9/8}\ ||\{ \cdot\cdot\cdot\}||_{1,8/7}\ ds,
     \nonumber\\
\int_{I_2}||E_1(t-s)\Pc\{ \cdot\cdot\cdot\}||_8\ ds &\le&
      C\int_{t/2}^{t-1} |t-s|^{-9/8}\ ||\{ \cdot\cdot\cdot\}||_{1,8/7}\ ds,
\nonumber\\
\int_{I_3}||E_1(t-s)\Pc\{ \cdot\cdot\cdot\}||_8\ ds &\le&
      C\int_{t-1}^t |t-s|^{-3/8}\ ||\{ \cdot\cdot\cdot\}||_{1,8/7}\ ds,
\label{eq:eta3b}
\ea
where
\be ||\{ \cdot\cdot\cdot\}||_{1,8/7}\le\
C\left(|A|^2\ ||\varphi^2\eta||_{1,8/7}\ +\
               |A|\ ||\varphi\eta^2||_{1,8/7}\ +\
              ||\eta^3||_{1,8/7}\right).
\ee
\medskip
\begin{prop}
\ba 
 &&(i)\ |A|^2\ ||\varphi^2\eta||_{1,8/7}\ \le \ C_\varphi\ |A|^2\
\left(||\eta||_8\ +\ ||B\eta||_4\right),\nonumber\\
 &&(ii)\ |A|\ ||\varphi\eta^2||_{1,8/7}\ \le\ 
 C_\varphi\ |A|\ \left(||\eta||_8^2\ + \ ||B\eta||_4\
||\eta||_8\right),\nonumber\\
&&(iii)\ ||\eta^3||_{1,8/7}\ \le\ C\left(||\eta||_{2,1},\varphi\right)\
||\eta||_8^{5/3}
\label{eq:eta3c}
\ea
\end{prop}
\bigskip

\nit{\it proof:} We prove part (i). The estimates (ii) and (iii) follows
similarly.
\ba ||\varphi^2\eta||_{1,8/7}\ &\le&\ 
              C\left(||\varphi^2\eta||_{8/7}\ +\ 
              ||\varphi\D\varphi\eta||_{8/7}\ +\ 
              ||\varphi^2\D\eta||_{8/7}\right)\nn \\
              &\le&\ 
              C\left(||\varphi^2||_{4/3}\ ||\eta||_8\ +\ 
              ||\varphi\D\varphi||_{4/3}\ ||\eta||_8\ +\
              ||\varphi^2||_{8/5}\ ||\D\eta||_4\right),
			  \label{eq:db0}
\ea
by H\"older's inequality. 
% Lp boundedness stuff
%
To express the right hand side of (\ref{eq:db0}) in terms of $||B\eta||_4$,
and thereby completing the proof of part (i), it suffices to show that:
\be
||\D\eta||_4\ \le\  C ||B\eta||_4.\nn\ee
This follows if we show that the operator 
$\D_iB^{-1}$ is bounded on $L^p$ (with $p=4$). 
\bigskip

To prove the $L^p$ boundedness of $\D_iB^{-1}$, we can apply the results
on the wave operator, $W_*$ in \S2.2. Indeed, for any $g\in L^p$, we
have using the boundedness of the wave operators on $W^{1,p}$, for $p\ge1$,
that
\ba
\|\D_iB^{-1}g\|_p\ &=&\ \|\D_i W_+ B_0^{-1}W_+^*g\|_p\nn\\
                   &\le&\ C\| W_+ B_0^{-1}W_+^*g\|_{W^{1,p}}\nn\\
				   &\le&\ C\|B_0^{-1}W_+^*g\|_{W^{1,p}}\nn\\
				   &\le&\ C\|W_+^*g\|_p\ \le\ C\|g\|_p.
\nn\ea

\nit{\bf Remark:} We offer here an alternative proof which does not make
use of the wave operators, and therefore applies under weaker hypotheses
on $V$. 
We begin with the 
square root formula see \cite{kn:RS1} and (\ref{eq:Katosqrt1}):

\nit For any $\psi\in{\cal D}(B^2)$:
\be
B^{-1}\ =\ \pi^{-1}\ \int_0^\infty\ w^{-1/2}(B^2 + w)^{-1}\ dw.
\label{eq:Katosqrt}
\ee
By (\ref{eq:Katosqrt}) and the second resolvent formula we have:
\be
\D_i B^{-1}\ =\  \D_i B_0^{-1} 
		\ +\ \pi^{-1}\ \int_0^\infty\ w^{-1/2}\ \D_i (B_0^2 + w)^{-1}\ V\ 
		 (B^2 + w)^{-1}\ dw.
\label{eq:Kato2}
\ee
The boundedness in $L^p$  ($p\ge1$) of the operator $\D_i B_0^{-1}$ holds because 
$\xi_i(|\xi|^2 +m^2)^{-1/2}$ is a multiplier on $L^p$; see \cite{kn:Stein}.
Similarly, $\D_i (B_0^2 + w)^{-1}$ is bounded on $L^p$ for any $w>0$ and
estimation of the second term in (\ref{eq:Kato2}) is reduced to estimation
of the norm of the operator $V\ (B^2 + w)^{-1}$. 
 Note that:
\be
(B^2 + w)^{-1}\ =\ \int_0^\infty e^{-t(B^2+w)}\ dt\label{eq:heatrep}
\ee
Since $\inf\sigma(B^2)=\Omega^2>0$,
\be 
\| e^{-t(B^2-\Omega^2/2)} \|_{{\cal B}(L^p)}\ \le\ C\nn
\ee 
and therefore
\be
\|(B^2 + w)^{-1}\|_{ {\cal B}(L^p)}\  \le\ C\ \int_0^\infty e^{-t(\Omega^2/2 +w)}
\ \le\ C'(1+w)^{-1}.\nn\ee
Boundedness in $L^p$ of $\D_i B^{-1}$ now follows. Namely,
\be
\|\D_i B^{-1}\|_{{\cal B}(L^p)}\ \le\ 
\|\D_i B_0^{-1}\|_{{\cal B}(L^p)}\
 +\ C'\|\D_i (B_0^2 + w)^{-1}\|_{{\cal B}(L^p)}\ \|V\|_\infty\ \int_0^\infty
  w^{-1/2}(1+w)^{-1}\ dw\ < \infty.\nn
  \ee
\medskip

From (\ref{eq:eta3c}) we see that $||B\eta||_4$ must be estimated.
Recall that
\be B\eta\ =\ B\eta_1\ +\ B\eta_2\ +\ B\eta_3.\ee
By (\ref{eq:eta1})
\be B\eta_1\ =\ E_0(t)B\Pc u_0\ +\ E_1(t)B\Pc u_1,\ee
and so 
\be ||B\eta_1(t)||_4\ \le\ C\jt^{-1/2}\ \|\bu_0\|_{\bX};
 \label{eq:beta1}\ee   
 see (\ref{eq:bX}) for definition of  $\|\bu_0\|_{\bX}$.

By (\ref{eq:eta2})
\be 
B\eta_2\ =\ \lambda\int_0^t E_1(t-s)\ a^3(s)\ B \Pc\varphi^3\ ds,
\ee
which can by Theorem 2.3 be estimated as                     
\be
||B\eta_2(t)||_4\ \le\ C|\lambda|\int_0^t|t-s|^{-3/4}\ |A(s)|^3\
 \|\Pc\varphi^3\|_{4/3}\ ds\ \le \ C_\varphi |\lambda|\ [A]_{1/4}^3\ 
                                   \jt^{-1/2}. 
\label{eq:Beta2}
\ee
The estimation of $||B\eta_3(t)||_4$ is more involved. By
(\ref{eq:eta3}),
\be
B\eta_3\ =\ \lambda\int_0^t\ \sin B(t-s)\ \Pc
   \left[ 3a^2\ \varphi^2\ \eta\ +\ 3a\ \varphi\eta^2\ +\ \eta^3\
   \right]\ ds.
\ee
By Theorem 2.3, 
\be
||B\eta_3(t)||_4\ \le\ C|\lambda|\int_0^t|t-s|^{-1/2}
              \left[\ |A|^2\ ||\varphi^2\eta||_{1,4/3}\ +\
                       |A|\ ||\varphi\eta^2||_{1,4/3}\ +\
              ||\eta^3||_{1,4/3}\ \right]\ ds.\label{eq:beta3}
\ee
The following proposition provides estimates for the integrand. It is
proved in the same manner as Proposition 7.2.

\begin{prop}
\ba 
 |A|^2\ ||\varphi^2\eta ||_{1,4/3}\ &\le&\ 
C_\varphi\ |A|^2\ \left( ||\eta ||_8\ +\ ||B\eta_1 ||_4\ +\ ||B\eta_2 ||_4\
+\ ||B\eta_3 ||_4\right) , \nonumber\\
 |A|\ ||\varphi\eta^2 ||_{1,4/3}\ &\le&\  
C_\varphi\ |A|\ \left( ||\eta ||_8^2\ +\ ||\eta ||_8\  
||B\eta ||_4\ \right) 
  \nonumber\\
 ||\eta^3 ||_{1,4/3}\ &\le&\  3\ ||\eta ||_8^2\ ||\D\eta ||_2.
\ea
\end{prop}
\bigskip

Anticipating the behavior 
$$|A(t)|\sim t^{-1/4},\ ||\eta(t)||_8\sim\ t^{-3/4},$$
and using (\ref{eq:beta1}), we find that $B\eta_3$
% and (\ref{eq:beta2)???
is driven by terms which are formally of order $\jt^{-1}$. 
Estimation in 
$L^4$ will lead to convolution of $\jt^{-1}$ with
$\jt^{-1/2}$ giving a rate of $\jt^{-1/2 + \delta_0}$,
for any $\delta_0>0$. 

\bigskip

For $0\le t\le T$, with $T$ fixed and arbitrary, we have:

\ba 
|A|^2\ ||\varphi^2\eta||_{4/3,1}\ &\le&\
C_\varphi\ [A]_{1/4}^2\ \left([\eta ]_{8,3/4}\jt^{-5/4}\
 +\ [B\eta_1]_{4,1/2}\jt^{-1}\right.\nn\\
 &+&\ \left. 
    [B\eta_2]_{4,1/2}\jt^{-1}\ +\ 
    [B\eta_3]_{4,1/2-\delta_0} \jt^{-1+\delta_0}\right),
\nn
\ea
where $\delta_0$ is positive and arbitrary.
Also,
\ba
|A|\ ||\varphi\eta^2||_{1,4/3}\ &\le&
 C_\varphi\ [A]_{1/4}\left(\ [\eta ]_{8,3/4}^2\jt^{-7/4}\
   +\ [\eta ]_{8,3/4}[B\eta_1]_{4,0}\jt^{-1}\ \right)
\nonumber\\
||\eta^3||_{1,4/3}\ &\le&\ ||\eta ||_{1,2}\ [\eta ]_{8,3/4}^2\ 
 \jt^{-3/2}.
\nonumber
\ea
It follows that for $0\le t\le T$
\be
||B\eta_3 ||_4\ \le\ C_\varphi |\lambda|\
 \int_0^t\ |t-s|^{-1/2}\ ds \jt^{-1+\delta_0}\ 
G(A,\eta,T),\nonumber
\ee
where
\ba
G(A,\eta,T)\ &\equiv&\ 
[A]_{1/4}^2\ \left(\ [\eta ]_{8,3/4}\ +\ [B\eta_1]_{4,1/2}\ +\
             [B\eta_2]_{4,1/2}\ +\ [B\eta_3]_{4,1/2-\delta_0}\
\right)\nonumber\\
 &+& [A]_{1/4}\ \left(\ [\eta ]_{8,3/4}^2\ +\ [\eta ]_{8,3/4}\
[B\eta_1]_{4,0}\ \right)\ + ||\eta ||_{1,2}\ [\eta ]_{8,3/4}^2.
\nonumber
\ea

Therefore, for $0\le t\le T$:
\be
||B\eta_3(t) ||_4\ \le\ C_\varphi |\lambda|\ G(A,\eta,T)\ 
\jt^{-1/2 +\delta_0},\
\label{eq:b3est}
\ee
for any $\delta_0>0$.
We now use the estimate (\ref{eq:b3est}) in (\ref{eq:eta3c}) in order to
obtain a bound on $||\eta_3(t) ||_8$. First, a consequence of
 (\ref{eq:eta3c}), (\ref{eq:beta1}), (\ref{eq:Beta2}) and
 (\ref{eq:b3est}) is that  for $0\le t\le T$:
\ba
|A|^2\ ||\varphi^2\eta||_{1,8/7}\ &\le&\ 
 C_\varphi\ [A]_{1/4}^2\ \left(\ [\eta ]_{8,3/4}\ +\
\|\bu_0\|_{\bX}\ +\ [A]_{1/4}^3\ +\ [B\eta_3 ]_{4,1/2-\delta_0}\
\right) \ \jt^{-1+\delta_0} ,\nonumber\\
|A|\ ||\varphi\eta^2||_{1,8/7}\ &\le&\ 
C_\varphi\ [A]_{1/4}\ \left(\ [\eta ]_{8,3/4}^2\ +\ 
 \{\|\bu_0\|_{\bX}\ +\ [B\eta_3]_{4,1/2-\delta_0}\ 
+\ [A]_{1/4}^3 \}\ [\eta ]_{8,3/4}\ \right) \ 
\jt^{-1+\delta_0},\nonumber\\
||\eta^3||_{1,8/7}\ &\le&\ C_\varphi\left( ||\eta||_{1,2}\right) \ 
 [\eta ]_{8,3/4}^{5/3}\ \jt^{-5/4}.
\label{eq:b3estA}
\ea
Substitution of (\ref{eq:b3estA}) into  (\ref{eq:eta3a}-\ref{eq:eta3b})
 leads to an estimate for $||\eta_3(t) ||_8$.
For $0\le t\le T$:
\ba
||\eta_3(t) ||_8\ &\le&\ |\lambda|\ C_\varphi\ \jt^{-1+\delta_0}\ 
		  \left( \right.
               [A]_{1/4}^2  
               \{   
                [\eta ]_{8,3/4}\ +\ \|\bu_0\|_{\bX} +\ [A]_{1/4}^3\
                +\  [B\eta_3 ]_{4,1/2-\delta_0} 
               \} \nonumber\\
              &+&  [A]_{1/4}\ 
                \{\  [\eta ]_{8,3/4}^2\ +\ \left[\ 
				 \|\bu_0\|_{\bX} +\
                   [B\eta_3]_{4,1/2-\delta_0}\ +\ [A]_{1/4}^3\ \right]\
                      [\eta ]_{8,3/4}\ 
                \}\nonumber\\
              &+& \ C\left(||\eta||_{1,2},\varphi\right)\ 
                [\eta ]_{8,3/4}^{5/3}\ \left. \right), 
\label{eq:eta3A}
\ea
where $\delta_0$ is arbitary.

Finally we can now estimate $F_2(a,\eta_3)$ using 
 (\ref{eq:F1234}) and  (\ref{eq:eta3A}). Choosing $\delta_0$ so that  
$$-1+\delta_0\ =\ -3/4 -\sigma_0,\ {\rm  with}\  \sigma_0>0,$$
we have
\begin{prop}
\ba
 \left|  F_2(a,\eta_3)\right|\
 \ &\le&\ C_\varphi |\lambda |\ [A]_{1/4}^2\ 
 \left(\ [A]_{1/4}^2\{\ [\eta ]_{8,3/4}\ +\ \|\bu_0|_{\bX}\ +\ 
 [A]_{1/4}^3\ +\ [B\eta_3 ]_{4,1/4+\sigma_0}\ \}\right. \nonumber\\
 &+&  [A]_{1/4}\ \{\ [\eta ]_{8,3/4}^2\ +\ 
  \left[ \|\bu_0\|_{\bX}\ +\ 
[A]_{1/4}^3\ +\ [B\eta_3 ]_{4,1/4+\sigma_0}\ \right]\ [\eta ]_{8,3/4}
\}\ \nn\\
  &+&\left.\  \ C\left(||\eta||_{1,2},\varphi\right)\
                [\eta ]_{8,3/4}^{5/3}\ \right)\ 
                             \jt^{-5/4-\sigma_0}.  
\label{eq:eta3B}
\ea
\end{prop}
\bigskip

\centerline{\it Estimation of $ F_2(a,\eta_*^{nr1})$ and 
 $F_2(a,\eta_*^{nr2})$ }
\bigskip

By (\ref{eq:eta-2eps1})
\ba
\eta_*^{nr1}\ &=&\ -{\lambda\over2} A_0^3\ B^{-1}(B-3\Omega +i0)^{-1}\
e^{iBt}\ \Pc\ \varphi^3,\ {\rm and}\ \nonumber\\
\eta_*^{nr2}\ &=& -{3\lambda\over2}\ B^{-1}(B-3\Omega +i0)^{-1}\ 
 \int_0^te^{iB(t-s)}\ e^{i3\Omega s}\ A^2A'\ ds \Pc\varphi^3
\nonumber
\ea

Estimation using Proposition 2.2 gives:
\ba
\left|F_2(a,\eta_*^{nr1})\right|\ &=&\ 3\ |\lambda|\ |a|^2\
\left|\int\varphi^3\eta_*^{nr1}\right|\nonumber\\
 &=&\ C|\lambda|^2\ |A_0|^3 |A|^2\ 
 \left|\ \int\ \varphi^3\ \left[ B(B-3\Omega +i0 )\right]^{-1}\
           e^{iBt}\ \Pc\ \varphi^3\ \right| \nonumber\\
 &\le&\ C|\lambda|^2\ |A_0|^3 |A|^2\ \|\
\la x\ra^\sigma\varphi^3\ \|_2\  \|\la x\ra^{-\sigma}
 (B-3\Omega +i0)^{-1}\ e^{iBt}\ 
 \Pc B^{-1} \varphi^3\|_2 \nonumber\\
 &\le&\ C_{\varphi}\   |\lambda|^2\ |A_0|^2\ 
   [A]_{1/4}^2\ \jt^{-{5\over4} +\delta}.
\ea 
Here, we have used that 
$1/2 + 6/5 = 5/4 + \delta$, for some $\delta >0$.
The constant $C_\varphi$ depends on 
$ \|\la x\ra^\sigma B^{-1} \la x \ra^{-\sigma}\|_{{\cal B}(L^2)}$.  
It is easy to check that this norm is bounded if we replace $B^{-1}$ by
$B^{-2}$. The estimate of interest is reduced to this case using the
Kato square root formula (\ref{eq:Katosqrt}).
Similarly, we have 
\ba
\left|  F_2(a,\eta_*^{nr2})\right|  &=&\ 3|\lambda |\  |a|^2\ 
  \left| \int\varphi^3\eta_*^{nr2}\right| \nonumber\\
 &\le&\ C\ |\lambda|^2\ |A|^2\ \left|\ \int\ \varphi^3 \left[
B(B-3\Omega +i\varepsilon)\right]^{-1}\ \int_0^t\
e^{i(t-s)(B-3\Omega)}\ A^2 A'\ ds\ \Pc \varphi^3\ \right|\nn\\
&\le&\ C_\varphi\ |\lambda |^2\ |A|^2\ 
 \int_0^t\la t-s\ra^{-{3\over2}+\mu}\ 
|A|^2\ |A'|\ ds\ \nn\\
&\le&\  C_\varphi\ |\lambda |^2\ [A]_{1/4}^2\ [|A|^2|A'|]_{5/4}\ \la
t\ra^{-{7\over4}},  
\ea
where we have taken $\mu$ sufficiently small 
 so that the last integral $t-$ integral is
convolution with an $L^1$ function, which then preserves the decay rate
 , $\la t\ra^{-{7\over4}}$, 
which exceeds $\la t\ra^{-{5\over4}}$
\bigskip

\centerline{\it Estimation of $E_{2j}^{nr}$:}
\bigskip

 To estimate (\ref{eq:4.40}) we use Propositions
 2.1 and 2.2. Due to the singularity in the resolvent at frequency
 $3\Omega$, the case  $j=7$, is most difficult and we focus on it. 
  To treat this singularity, we use
 Proposition 2.2, for $n=3$:
 \be
 \| \la x\ra^{-\sigma}(B-\zeta_7)^{-2}e^{iB(t-s)}\la x\ra^{-\sigma}\psi
 \|_2\ \le\ C\la t-s\ra^{-{6\over5}}\|\psi\|_{1,2}. \nn
 \ee
Use of this estimate in (\ref{eq:4.40}) yields 
\ba
&& |E_{2j}^{nr}|\ \le\nn\\
 &&\ \ \ \ C_\varphi |\lambda |^2
\ \left(\ [|A|^4|A'|]_{7/4} \jt^{-7/4}\ +\ [A]_{1/4}^2\jt^{-13/8}\ +\ 
    [|A|^4|A''| + |A|^3|A'|^2]_{9/4}\jt^{-13/8}\ \right),
\nn
\ea
 from which estimate (vii) of Proposition 7.1 follows. This finally
completes the proof of Proposition 7.1.

Proposition 7.1 can now be used to obtain the desired estimate for 
${\bf E}(t)$ and therefore ${\bf\tilde A}(t)$, for $|A|$ sufficiently
small. Note that the right hand side of the estimates depend on 
$|A'|$ and $|A''|$. These can each be estimated directly from the equation 
 for $A$, (\ref{eq:4.41}), and its derivative. The result is that 
$|A'|$ and $|A''|$  may be estimated by $[A]_{1/4}^3\la t\ra^{-{3\over4}}$,
the only effect being a change in the multiplicative constants which is
independent of $A$ and $\eta$. This observation together with
Proposition 7.1 yields:

%HELP
\begin{prop}
There is positive number $\delta$ such that 
\be
\left| {\bf E}(t)\right| \le\ Q_0(A,\eta) 
 \la t\ra^{-{5\over4}-\delta}
\label{eq:EQbound}
\ee
where 
\be
 Q_0(A,\eta)\ =\  
 [\eta ]_{8,3/4}^3\ +\ [A]_{1/4}\ 
 \left(\ 1\ +\ C\left( [A]_{1/4},[\eta ]_{8,3/4},
       [B\eta_3]_{4,1/4+\sigma_0}, \|\bu_0\|_{\bX}\ 
                                                    \right)\ 
 \right) 
\label{eq:Q0}
\ee
\end{prop}
\medskip

It now remains to estimate $||\eta ||_8$. 

\ba
||\eta ||_8 &\le&\ ||\eta_1 ||_8\ +\ ||\eta_2 ||_8\ +\ ||\eta_3 ||_8\nonumber\\
    &\le&\ ||E_0(t) \Pc u_0||_8\ +\ ||E_1(t) \Pc u_1||_8\ +\ 
      |\lambda |\int_0^t ||E_1(t-s)a^3(s) \Pc\varphi^3||_8\ ds\ +\ 
         ||\eta_3 ||_8.
\nonumber
\ea
Using Theorem 2.3 and the bound (\ref{eq:eta3A}) we get
\begin{prop}
\ba
||\eta ||_8 &\le&\ C_\varphi\ \left(\ 
\|\bu_0\|_{\bX}\ +\ [A]_{1/4}^3\right.  \nonumber\\
 &+&\ [A]_{1/4}^2 \{ \ [\eta ]_{8,3/4}\ +\ \|\bu_0\|_{\bX}\ +\
 [A]_{1/4}^3\ + [B\eta_3]_{4,1/4+\sigma_0} \}\nonumber\\
 &+&\ [A]_{1/4}\ \{\ [\eta ]_{8,3/4}^2\ +\ 
                  \left[ \|\bu_0\|_{\bX}\ +\ [A]_{1/4}^3\ +\
                   [B\eta_3]_{4,1/4+\sigma_0}  \right]\ [\eta ]_{8,3/4}\}
  \nonumber\\
 &+&\ \left.  C\left( ||\eta ||_{1,2},\varphi\right)\ [\eta ]_{8,3/4}^{5/3}\
\right)\ \la t\ra^{-3/4} . 
\label{eq:eta3C}
\ea
\end{prop}
\bigskip
The following proposition summarizes our labors.

\begin{prop}
For  any $T>0$:
\ba
[\eta ]_{8,3/4}(T) 
 &\le&\ C_\varphi \left(\ \|\bu_0\|_{\bX}\ +\
   [A]_{1/4}^3(T)\ +\ [B\eta_3 ]_{4,1/4+\sigma_0}^2(T)\right. \nn\\
&+& \left.  
    C_\varphi\left( ||\eta ||_{1,2}\right)\ [\eta ]_{8,3/4}^{5/3}(T)\ 
      \right) \nonumber\\
\ [A]_{1/4}^4(T) &\le&\ \left( |A_0|^4 + Q_0(A,\eta)^{8\over5}\ \right)
\nonumber\\
\ [B\eta_3]_{4,1/4+\sigma_0}(T) &\le&\ 
 \left( [A]_{1/4}^2(T)\ +\ [\eta ]_{8,3/4}^2(T)\ +\ [B\eta_2]_{4,1/2}^2(T)\ 
  +\ [B\eta_3]_{4,1/4+\sigma_0}^2(T)\ \right),\nonumber\\
\sup_{0\le t\le T}||\eta(t)||_{1,2}\ &\le&\ C\ {\cal E}(u_0,u_1)\le C
        \|u_0 , u_1\|_{1,2}.
\nonumber
\ea
\end{prop}

The first three estimates are proved above while the last follows from
conservation of energy (see section 1) and the decomposition of the
solution.

Now define
\be M(T)\ \equiv\  [\eta ]_{8,3/4}(T)\ +\ [A]_{1/4}(T)\ +\
[B\eta_3]_{4,1/4+\sigma_0}(T).\nonumber
\ee
Then, combining the estimates of the previous proposition we have, for
some $\alpha >0$:
\be
M(T) \left(\ 1-M(T)^\alpha\ \right)\ \le\ C_\varphi\ 
\|\bu_0\|_{\bX}.
\nonumber
\ee

If $M(0)$ and $\|\bu_0\|_{\bX}$ 
are sufficiently small, we have by the continuity of $M(T)$,   
that there is a constant $M_*$, which
is independent of $T$, such
that for all $T$ 
\be M(T)\le M_*.
\label{eq:aprioriest}
\ee
This completes the proof of Theorem 1.1.

\bigskip
\section{ Summary and discussion}
We have considered a class of nonlinear Klein-Gordon equations,  
 (\ref{eq:nlkg}),
 which are perturbations of a linear dispersive equation which has a
time periodic and spatially localized (bound state) solution. The
unperturbed and perturbed dynamical systems are Hamiltonian. We have shown
that if a  
 nonlinear resonance condition (\ref{eq:nlfgr}) holds,     
  then solutions with
sufficiently small initial data tend to zero as $t\to\pm\infty$. This
resonance condition is a nonlinear variant  of the {\it Fermi golden rule}
 (\ref{eq:fgr}). A consequence of our result is that  
 time-periodic and spatially localized solutions do not persist under
small Hamiltonian perturbations. Time-decay of small
amplitude solutions  
is also a  property of the translation invariant ($V\equiv0$) 
 nonlinear Klein-Gordon equation. However, the presence of a bound
state of the unperturbed problem, 
  causes  a  nonlinear resonance leading to the anomalously slow
 radiative decay of solutions. 

We now conclude with some further remarks on the results of this paper,
 mention 
 directions currently under
investigation and some open problems.

\bigskip

\nit {\bf 1.\ Anomalously slow time-decay rates:}
It is natural to compare the time-decay
 rate of solutions described by Theorem 1.1 with those
  of related problems.

   (1a) {\it Translation invariant linear Klein-Gordon equation}\
	($V\equiv
	 0$ and $\lambda=0$):

	  \nit We shall refer to {\it free
	   dispersive rates of decay} as those associated with the constant
		coefficient equation:
	 \be
  \D_t^2 u\ -\ \Delta u\ + \ m^2u\ =\ 0.\label{eq:freeKG}
   \ee
	Results on this are presented in \S2. Roughly speaking,
	  if the initial data has a sufficient number of
	derivatives in $L^p,\ 1 < p\le 2$, then the solution
	decays at a
	  rate
	${\cal O}(t^{-n({1\over2}-{1\over p'})})$ in
	$L^{p'}$. Here, $p^{-1} +  (p')^{-1}=1$.

	(1b) {\it Translation invariant nonlinear
		  Klein Gordon equation}\
		($V\equiv 0$ and $\lambda\ne0$):\

 \nit
 For small initial conditions it has
 been shown that solutions decay at  free dispersive rates;
     see \cite{kn:strauss} and references
		  cited therein.

	 (1c) {\it Linear Klein-Gordon equation with a potential having a
	 bound
	 state, as hypothesized}
	 \ ($V\ne0$ and $\lambda=0$):
	 \medskip
\nit By the spectral theorem, a typical solution will decompose into
a
linear superposition of (i) a bound state part, of the form:
$R_0\cos(\Omega t+\rho_0)\varphi(x)$, and (ii) a
 part which disperses to zero at free
 dispersive rates.

 (1d) {\it Nonlinear Klein-Gordon equation with a potential having a
 bound
 state, as hypothesized}\ ($V\ne0$ and $\lambda\ne0$):
  Whereas the decaying part of the solution in the above cases tends
  to
   zero at a
	free dispersive rate, Theorem 1.1 implies that the decay rate is
	  anomalously slow. In particular, the decay rate
	   obtained in $L^8$ is ${\cal O}(t^{-{1\over4}})$, while the
	   free
		dispersive rate in $L^8$ is ${\cal O}(t^{-{9\over8}})$.
		This slow rate of decay, due to the nonlinear resonant
		interactions
		 give rise to a
		 long-lived or {\it metastable states}.

\nit{\bf 2. Dispersive Hamiltonian normal form}

In a finite dimensional Hamiltonian system of one degree of freedom,
the
general normal form is:
\be A' = i\left(c_{10} + c_{21}|A|^2 + c_{32}|A|^4 +...+
		  c_{n+1,n}|A|^{2n}+...\right)A,
		  \ee
		  where $c_{n,n+1}$ are real numbers. In the present context we
		  have the
		  {\it dispersive  Hamiltonian normal form}
		  \be A' = \left(k_{10} + k_{21}|A|^2 + k_{32}|A|^4
		  +...+ k_{n+1,n}|A|^{2n}+...\right)A,
		  \ee
		  where
		  $$k_{n+1,n}=d_{n+1,n} + ic_{n+1,n}$$
		   are, in general,
		   numbers with real and imaginary part. Contributions to the
		   real parts of
		   coefficients come from resonances with the continuous
		   spectrum.
 If (\ref{eq:nlfgr}) fails 
 there are two possibilities; either (a) $3\Omega$ does not lie in the
 continuous spectrum of $B$ (it lies in the gap $(0,m)$), or (b) $3\Omega$
 lies in the continuous spectrum of $B$ but we are in the non-generic
 situation where the projection of $\varphi^3$ onto the generalized
 (continuum) eigenmode of frequency $3\Omega$ is zero. 
 In either case, we  expect that, typically, internal dissipation
  would arise in the normal form at higher order. 
  More precisely, we conjecture that the leading order nonzero
 $d_{n_*+1,n_*}$, which would correspond 
  to a resonance with the continuum of a
 higher harmonic: $q_*\Omega\in\sigma_{cont}(B)$, $q_*>3$ is always   
 negative. (If the sign of $d_{n_*+1,n_*}$  were positive,
	 this would seem to be in violation
	  of the conservation of energy and the implied Lyapunov stability
	   of the zero solution.)  
   The corresponding
  decay rate would then be slower, specifically  ${\cal
   O}( |t|^{-{1\over 2n_*}} )$.

From this perspective one may expect the existence of breather solutions
of integrable nonlinear flows like the sine-Gordon equation or the extremely
long-lived breather -like states of the $\phi^4$ - model as
corresponding to the case where to all orders 
 the coefficients  $d_{n+1,n}$ are zero. Does the vanishing of all such
 coefficients have an interpretation in terms of the infinitely many
 time-invariants for the integrable flow?

\nit {\bf 3. Multiple bound state problems:}  
 Of interest are problems 
  where the underlying potential, $V(x)$, supports more than
 one bound state. What is the large time behavior of such systems?  
 Our analysis and the above remarks suggest that 
  a more general normal form could be developed, and the
 corresponding ( in general slower) decay rates anticipated.
 Is it possible that a resonance among the multiple 
"discrete oscillators" can be arranged so that one gets a persistence of
nondecaying solutions, or does radiation to the continuum always win out?

\nit {\bf 4. Relation to center manifold theory:}
Our program of decomposing the original conservative 
 dynamical system into a finite
dimensional dynamical system (\ref{eq:dosc}) which is weakly coupled to
an infinite dimensional dynamical system is one commonly used in
dissipative
dynamical systems. There, it is often possible using the center   
manifold approach \cite{kn:Carr}, \cite{kn:VI}, to construct an invariant 
center-stable manifold. The dynamics in a neighborhood of an equilibrium
 point are characterized by an exponentially fast contraction on to the
 stable-center manifold. The contraction is exponential because the part
 of the linearized spectrum associated with  the infinite dimensional
 part of the dynamics is contained in the left half plane. In the
 current context of conservative dynamical systems, the linearization
 about the equilibrium point (here $u\equiv 0$) lies on the imaginary
 axis; in particular, two complex conjugate eigenvalues $\pm i\Omega$
 and continous spectrum from $\pm im$ to $\pm i\infty$.
  The analogue of dissipation is the mechanism of  dispersive radiation
  of energy, related to the continuous spectrum, and the associated
  algebraic time-decay. 
 
 There is recent work on the 
  application of center manifold methods to certain special
 conservative dynamical systems of nonlinear Schr\"odinger type; 
 see the center manifold analysis of \cite{kn:PW} applied to problem studied
 by the authors in  \cite{kn:SW1bs}.  
  It would be of interest to understand whether  
  geometric insight on the structure of the phase space for 
   problems of the type
 considered in this paper can be obtained using the ideas of  
   invariant manifold theory. 

\nit {\bf 5. Systems with disorder:}
In this paper, we have seen the effect of a single localized defect on
the wave propagation dynamics in a nonlinear system. 
Of great interest would be an understanding of the effects of a  
spatially random distribution  of defects modeled, for example, 
 by random potential  
 $V(x)$ on the localization of energy in nonlinear systems
such as (\ref{eq:nlkg}). Related questions are studied in \cite{kn:FSW},
 \cite{kn:GKSV}, \cite{kn:Bronski}.

\bigskip
%%%%%
%%%%%
% BIBLIOGRAPHY

%

\end{document}